\documentclass[11pt]{article}

\usepackage{amsmath}
\usepackage{amssymb}
\usepackage{amstext}
\usepackage{latexsym}
\usepackage{color}
\usepackage{graphicx}
\usepackage{epsf}
\usepackage{epsfig}
\usepackage{subfigure}
\usepackage{rotating}
\usepackage{curves}
\usepackage{epic}
\usepackage{amsthm}

\usepackage{amsfonts}
\usepackage[T1]{fontenc}
\usepackage[latin1]{inputenc}
\usepackage{authblk}
\usepackage{tikz}
\usetikzlibrary{shapes,arrows}
\usepackage{graphics}
\usepackage{verbatim}
\usepackage{color}
\usepackage{hyperref}

\newcommand{\be}{\begin{equation}}
\newcommand{\en}{\end{equation}}

\newcommand{\1}{1 \!\! 1}

\newtheorem{thm}{Theorem}
\newtheorem{cor}[thm]{Corollary}

\newtheorem{defi}{Definition}[section]
\newtheorem{lem}[defi]{Lemma}

\newtheorem{Theo}{Theorem}[section]

\newcommand{\bedefin}{\begin{defi}}
\newcommand{\findefi}{\end{defi} \medskip}

\newcommand{\betheo}{\begin{theorem}$\!\!${\bf \,\,\,}}
\newcommand{\entheo}{\end{theorem}}
\newcommand{\enth}{\end{theorem}}

\newcommand{\becor}{\begin{cor}$\!\!${\bf .}}
\newcommand{\encor}{\end{cor}}

\newcommand{\belem}{\begin{lem}$\!\!${\bf }}
\newcommand{\enlem}{\end{lem}}

\newcommand{\prf}{\noindent{\bf{ Proof}\,\,}}

\newcommand{\bea}{\begin{eqnarray}}
\newcommand{\ena}{\end{eqnarray}}

\newcommand{\beano}{\begin{eqnarray*}}
\newcommand{\enano}{\end{eqnarray*}}

\newcommand{\bee}{\begin{enumerate}}
\newcommand{\ene}{\end{enumerate}}

\newcommand{\bei}{\begin{itemize}}
\newcommand{\eni}{\end{itemize}}

\newcommand{\betab}{\begin{tabular}}
\newcommand{\entab}{\end{tabular}}

\newcommand{\bd}{\begin{displaymath}}

\newcommand{\h}{{\mathfrak H}}

\newcommand{\ba}{\mathbf a}

\newcommand{\bk}{\mathbf k}
\newcommand{\bp}{\mathbf p}
\newcommand{\bq}{\mathbf q}
\newcommand{\bs}{\mathbf s}
\newcommand{\bu}{\mathbf u}
\newcommand{\bx}{\mathbf x}

\newcommand{\bv}{\mathbf v}

\newcommand{\bQ}{\mathbf Q}
\newcommand{\bP}{\mathbf P}

\catcode `\@=11 \@addtoreset{equation}{section}

\catcode `\@=12

\begin{document}

\title{The Symmetry Groups of Noncommutative Quantum Mechanics and Coherent State Quantization}
\author{S. Hasibul Hassan Chowdhury$^\dag$ \\
  and S. Twareque Ali$^{\dag\dag}$\\
\medskip
\small{Department of Mathematics and Statistics,\\ Concordia University, Montr\'eal, Qu\'ebec, Canada H3G 1M8}\\

{\footnotesize $^\dag$e-mail: schowdhury@mathstat.concordia.ca\\
 $^{\dag\dag}$e-mail: stali@mathstat.concordia.ca}}

\date{\today}

\maketitle

\begin{abstract}
We explore the group theoretical underpinning of noncommutative quantum mechanics for a system
moving on the two-dimensional plane.  We show that the pertinent groups for the system are the
two-fold central extension of the Galilei group in $(2+1)$-space-time dimensions and the two-fold
extension of the group of translations of $\mathbb R^4$. This latter group is just the standard
Weyl-Heisenberg group of standard quantum mechanics with an additional central extension. We
also look at a further extension of this group and discuss its significance to noncommutative
quantum mechanics. We build unitary irreducible representations of these various groups and
construct the associated families of coherent states. A coherent state quantization of the underlying phase space is then carried out, which is shown to lead to exactly the same commutation relations as usually
postulated for this model of noncommutative quantum mechanics.

\end{abstract}

\section{Introduction}\label{sec:intro}
Noncommutative quantum mechanics is a much frequented topic of research these days. The
expectation here is that a  modification, or rather an extension, of standard quantum mechanics
is needed to model physical space-time at very short distances. In this paper we
restrict ourselves to
the version of non-commutative quantum mechanics which describes a quantum system with two
degrees of freedom and in which, in addition to having the usual canonical commutation relations,
one also imposes an additional non-commutativity between the two position coordinates, i.e.,
\be
  [Q_i , P_j ] = i\hbar \delta_{ij}I,\; \; i, j = 1,2, \;\; \qquad [Q_1, Q_2] = i\vartheta I,
\label{nc-comm-relns1}
\en
Here the $Q_i, P_j$ are the quantum mechanical position and momentum observables, respectively,
and $\vartheta$ is a (small, positive) parameter  which measures the additionally
introduced noncommutativity between the observables of the two spatial coordinates. The limit
$\vartheta = 0$ then corresponds to standard (two-dimensional) quantum mechanics.
The literature, even on this rather focused and simple model, is already extensive. We refer to
\cite{delducetalt,scholtzt} and the many references cited therein for a review of the background
and motivation behind the model.
One could continue with this game of increasing noncommutativity between the observables by
augmenting the above system by an additional commutator between the two momentum operators:
\be
  [P_i , P_j ] = i\gamma\delta_{ij} I\; , \qquad i,j = 1,2\; ,
\label{non-comm-relns2}
\en
where $\gamma$ is yet another positive parameter. Physically, such a commutator would signal,
for example, the presence of a magnetic field in the system \cite{delducetalt}.

  The purpose of this paper is two-fold. First, we undertake a group theoretical analysis of the
above sets of commutation relations, i.e., to find the groups behind noncommutative quantum
mechanics, in the same way as a centrally extended Galilei group \cite{levyt} or the
Weyl-Heisenberg group underlies
ordinary quantum mechanics. The second objective of this paper is to arrive at the commutation
relations (\ref{nc-comm-relns1}) by the method of coherent state quantization (see, for example,
\cite{aliengt} for a discussion of this method). This will involve
constructing appropriate families of coherent states, emanating from the groups underlying
noncommutative quantum mechanics, using standard techniques (see, for example, \cite{coherentt}).
It will turn out however, that the coherent states that we shall be using here are very different from the ones introduced in \cite{scholtzt}, in that ours come from the representations of
the related groups and satisfy standard resolutions of the identity condition.

\section{Noncommutative quantum mechanics in the two-plane and the (2+1)-Galilei group}

The (2+1)-Galilei group $G_\text{\tiny Gal}$ is a six-parameter Lie group. It is the
kinematical group of a classical, non-relativistic space-time having two spatial and one
time dimensions. It consists of  translations of time and space, rotations in the two dimensional
space and velocity boosts. As is well-known \cite{levyt}, non-relativistic quantum mechanics
can be seen as arising from representations of central extensions of the Galilei group.
We will thus be concerned here with the centrally extended (2+1)-Galilei group.
The Lie algebra
$\mathfrak G_\text{\tiny Gal}$ of the group $G_\text{\tiny Gal}$ has a three dimensional vector space of central extensions. This {\em extended algebra} has the
following Lie bracket structure (see, for example, \cite{bosecmpt,bosejmpt}),
\begin{eqnarray}\label{grpext}
[M,N_{i}]=\epsilon_{ij}N_{j} & \qquad & [M,P_{i}]=\epsilon_{ij}P_{j}\nonumber\\[2pt]
[H,P_{i}]=0 & \qquad &  [M,H]= \mathfrak h\nonumber\\[2pt]
[N_{i},N_{j}]=\epsilon_{ij} \mathfrak d & \qquad & [P_{i},P_{j}]=0\nonumber\\[2pt]
[N_{i},P_{j}]=\delta_{ij} \mathfrak m & \qquad &  [N_{i},H]=P_{i},
\end{eqnarray}
 ($i,j=1,2$ and $\epsilon_{ij}$ is the totally antisymmetric tensor with
 $\epsilon_{12}=-\epsilon_{21}$). The three central extensions are characterized by
 the three central generators $\mathfrak h , \mathfrak d$ and $\mathfrak m$ (they
 commute with each other and all the other generators). The $P_i$ generate space
 translations, $N_i$ velocity boosts, $H$ time translations and $M$ is the generator
 of angular momentum. Passing to the group level, the universal covering group
 $\widetilde{G}_\text{\tiny Gal}$, of $G_\text{\tiny Gal}$, has three central
 extensions, as expected. However, $G_\text{\tiny Gal}$ itself has only two central
 extensions (i.e., $\mathfrak h = 0$, identically \cite{bosecmpt}). We shall denote
 this 2-fold centrally extended $(2 + 1)$-Galilei group by  $G_\text{\tiny Gal}^\text{\tiny ext}$ and
 its Lie algebra by $\mathfrak G_\text{\tiny Gal}^\text{\tiny ext}$.

 A generic element of $G_\text{\tiny Gal}^\text{\tiny ext}$ may be written as
  $ g = (\theta, \phi, R , b, \mathbf v , \mathbf a) = (\theta, \phi, r)$, where $\theta, \phi \in \mathbb R$,
  are phase terms corresponding to the two central extensions, $b \in \mathbb R$
 a time-translation, $R$ is a $2\times 2$ rotation matrix, $\mathbf v \in \mathbb R^2$ a
 2-velocity boost,  $\mathbf a \in \mathbb R^2$ a 2-dimensional space translation and
 $r = (R, b, \bv , \ba)$.  The two
 central extensions are given by two cocycles, $\xi^1_m$ and $\xi^2_\lambda$, depending on
 the two real parameters $m$ and $\lambda$. Explicitly, these are,
 \bea
 \xi^1_m (r ; \; r') & = &
    e^{\frac {im}2 (\ba\cdot R\bv' - \bv\cdot R\ba' + b'\bv\cdot R\bv')}, \nonumber\\
 \xi^2_\lambda (r ; \; r') & = &
 e^{\frac {i\lambda}2 ( \bv \wedge R\bv')}\;,  \quad \text{where} \quad \bq\wedge \bp =
 q_1p_2 - q_2 p_1\; ,
 \label{cocycles}
\ena
($\bq = (q_1, q_2), \; \bp = (p_1 , p_2 )$).

 The
 group multiplication rule is given by
 \bea
  gg' & = & (\theta, \phi, R , b, \mathbf v , \mathbf a)(\theta^\prime,
 \phi^\prime, R' , b', \mathbf v^\prime , \mathbf a^\prime)\nonumber\\
   & = & (\theta + \theta' + \xi^1_m (r; \; r'), \;
      \phi + \phi' + \xi^2_\lambda(r; \; r'), \nonumber\\
      & \quad &  RR', b+b', \bv + R\bv',
      \ba + R\ba' + \bv b' )\; .
 \label{group-comp-law}
 \ena

The projective unitary irreducible representations (PURs) of $G_\text{\tiny Gal}^\text{\tiny ext}$,
from which
we can obtain its unitary irreducible representations,  have all
been computed in, e.g., \cite{bosejmpt}. In this paper we shall only consider the case where $m \neq 0$ and $\lambda \neq 0$.  These representations, realized on the Hilbert space
$L^2 (\mathbb R^2 , d\bk )$ (see (\ref{unitaryirrerep}) below), are characterized by  ordered pairs
$(m,\vartheta)$ of reals and by the number $s$, expressed as an integral multiple of
$\dfrac{\hbar}{2}$.
Here, $m$ is to be interpreted as the mass of the nonrelativistic system under study, while
$\lambda$ will be seen to be related to the parameter $\vartheta$ appearing in
(\ref{nc-comm-relns1}). The quantity
$s$ is the eigenvalue of the intrinsic angular momentum opearator $S$ (representating rotations
in the rest-frame). The physical significance of these quantities have been studied
 extensively in \cite{bosecmpt,bosejmpt,levyt} and \cite{Shudmukt}.

 Recall that we should take $\mathfrak h=0$ in (\ref{grpext}), to get the Lie algebra
 $\mathfrak G_\text{\tiny Gal}^\text{\tiny ext}$. In the representation Hilbert space of
the PURs of the group $G_\text{\tiny Gal}^\text{\tiny ext}$,
the basis elements of the algebra are realized as self-adjoint operators,
the two central elements appearing as multiples of the identity operator. Thus, the  operator
representation of $\mathfrak G_\text{\tiny Gal}^\text{\tiny ext}$ looks like
\begin{eqnarray}\label{algbrrepspc}
[\hat{M},\hat{N}_{i}]=i\epsilon_{ij}\hat{N}_{j}& \qquad & [\hat{M},\hat{P}_{i}]=
i\epsilon_{ij}\hat{P}_{j}\nonumber\\[2pt]
[\hat{H},\hat{P}_{i}]=0 & \qquad & [\hat{M},\hat{H}]=0\nonumber\\[2pt]
[\hat{N}_{i},\hat{N}_{j}]=i\epsilon_{ij}\lambda\hat{I} & \qquad & [\hat{P}_{i},\hat{P}_{j}]=0\nonumber\\[2pt]
[\hat{N}_{i},\hat{P}_{j}]=i\delta_{ij}m\hat{I} & \qquad & [\hat{N}_{i},\hat{H}]=i\hat{P}_{i}\; .
\end{eqnarray}
Here the operators $N_i$ generate velocity shifts.
The other operators $\hat{P}_{i}$, $\hat{M}$,
$\hat{H}$, and $\hat{I}$ are just the linear momentum, angular momentum, energy and the identity
operators, respectively, acting on $L^{2}(\mathbb{R}^{2},d\bk)$, the representation space
of the PUIRs of $G_\text{\tiny Gal}^\text{\tiny ext}$.

Consider next the so-called two-dimensional {\em noncommutative Weyl-Heisenberg group\/}, or the
group of {\em noncommutative quantum mechanics\/}. The group generators are the operators $Q_i, P_j$ and $I$, obeying the commutation relations (\ref{nc-comm-relns1}). The resulting algebra of operators
is also referred to as the the {\em noncommutative two-oscillator algebra\/.}
Realized on the Hilbert  space $L^{2}(\mathbb{R}^{2},d\bx)$ (coordinate representation) these
 operators can be brought into the form
\begin{eqnarray}\label{opsoscalgbr}
\tilde{Q}_{1}=x+\frac{i\vartheta}{2}\frac{\partial}{\partial y} & \qquad &
\tilde{Q}_{2}=y-\frac{i\vartheta}{2}\frac{\partial}{\partial x}\nonumber\\[2pt]
\tilde{P}_1=-i\hbar\frac{\partial}{\partial x} & \qquad &
\tilde{P}_2=-i\hbar\frac{\partial}{\partial y}.
\end{eqnarray}
If we add to this set the the Hamiltonian (corresponding to a mass $m$)
\begin{equation}\label{engyopncosc}
\tilde{H}=-\frac{\hbar^{2}}{2m}{\nabla}^{2}=-\frac{\hbar^{2}}{2m}
(\frac{\partial^2}{\partial x^{2}}+\frac{\partial^{2}}{\partial y^{2}}),
\end{equation}
the angular momentum operator,
\begin{equation}\label{angmomt}
\tilde{M}=-i\hbar\left(x\frac{\partial}{\partial x}-y\frac{\partial}{\partial x}\right).
\end{equation}
and furthermore, define $\tilde{N}_i = m\tilde{Q}_i, \; i =1,2$, then the resulting set of
seven operators is easily seen to obey the commutation relations
\begin{eqnarray}\label{liealgbrA}
[\tilde{M},\tilde{N}_{i}]=i\hbar\epsilon_{ij}\tilde{N}_{j} & \qquad &
[\tilde{M},\tilde{P}_i]=i\hbar\epsilon_{ij}\tilde{P}_j\nonumber\\[2pt]
[\tilde{H},\tilde{P}_{i}]=0 & \qquad &
[\tilde{M},\tilde{H}]=0\nonumber\\[2pt]
[\tilde{N}_{i},\tilde{N}_{j}]=i\epsilon_{ij}m^2\vartheta\tilde{I} & \qquad &
[\tilde{P}_{i},\tilde{P}_j]=0\nonumber\\[2pt]
[\tilde{N}_{i},\tilde{P}_j]=i\hbar\delta_{ij}m\tilde{I} & \qquad &
[\tilde{N}_{i},\tilde{H}]=i\hbar\tilde{P}_i,
\end{eqnarray}
Taking $\hbar = 1$ and writing $\lambda = m^2\vartheta$ this becomes exactly the same set of commutation relations as that in
(\ref{algbrrepspc}) of the Lie algebra $\mathfrak G_\text{\tiny Gal}^\text{\tiny ext}$. of the extended Galilei group. This tells us that the kinematical group of non-relativistic,
noncommutative quantum mechanics is the (2+1)-Galilei $G_\text{\tiny Gal}^\text{\tiny ext}$,
with two extensions, a fact which has already been noted and exploited in \cite{horduvsticht}.

 At this point we note that in terms of  $Q_{1}, Q_{2}$ and $P_{1}, P_{2}$,
the usual quantum mechanical position and momentum operators defined on
$L^{2}(\mathbb{R}^{2},dx\;dy)$, the noncommutative position operators
$\hat{Q}_{i}$ can be written as
\begin{eqnarray}
\hat{Q}_{1}&=&Q_{1}-\frac{\vartheta}{2\hbar}P_{2}\nonumber\\
\hat{Q}_{2}&=&Q_{2}+\frac{\vartheta}{2\hbar}P_{1}\label{trnsfrmsecnd}.
\end{eqnarray}
The above transformation is  linear and invertible and may be thought of as giving
a non-canonical transformation on the underlying phase space. Since $\hat{Q}_{i}=
Q_{i}\Leftrightarrow\vartheta=0$, the noncommutativity of the two-plane is lost
 if the parameter $\vartheta$ is turned off.
However, from the group theoretical discussion above we see that the noncommutativity of the two spatial coordinates should not just be looked upon as a result of this  non-canonical transformation. Rather, it is also the two-fold central extension of the (2+1)-Galilei group, governing nonrelativistic mechanics, which is responsible for it. The  extent to which the two spatial coordinates fail to be
commutative is encoded in of the representation parameters of the underlying group, namely,
$\vartheta$. It is noteworthy in this context that had we centrally extended the (2+1)-Galilei group
using only the cocycle $\xi_m^1$ in (\ref{cocycles}) (i.e., set $\vartheta$=0), we would have just obtained standard quantum mechanics. In this sense we claim that the group underlying noncommutative quantum mechanics, as governed by the commutation relations (\ref{opsoscalgbr}), is the doubly
 centrally extended $(2+1)$-Galilei group. (It is also worth mentioning in this context that
the noncommuting position operators $\hat{Q}_{i}$, arising from the (2+1)-centrally extended Galilei
group, also describe the position of the center of mass of the underlying non-relativistic
system.) (see \cite{Shudmukt}).

\section{Quantization using coherent states associated to non-commutative quantum mechanics}\label{CSquant}
In this section we first write down the unitary irreducible representations of the extended
Galilei group $G_\text{\tiny Gal}^\text{\tiny ext}$. Next we construct coherent states for
these representations, which we identify as being the
{\em coherent states of noncommutative quantum mechanics\/.} We then  carry out
a quantization of the underlying phase space using these coherent states, obtaining thereby the
operators $Q_i, P_i$  (see  (\ref{nc-comm-relns1})) of non-commutative quantum mechanics.
In the literature other coherent states have been defined for noncommutative quantum mechanics -- see, for example,
\cite{scholtzt}. These latter coherent states are basically the one-dimensional projection operators,
$\vert z \rangle\langle z \vert, \; z \in \mathbb C$, where $\vert z\rangle$ is the
well-known canonical coherent state, familiar from quantum mechanics (see, for example,
\cite{coherentt}). These coherent states have been shown to satisfy a sort of
an ``operator resolution
of the identity'' and have been used to study localization properties of systems obeying
noncommutative quantum mechanics. By contrast, the coherent states which we obtain
(see (\ref{coherentsttgalilei}) below),
using the representations of the group $G_\text{\tiny Gal}^\text{\tiny ext}$, i.e., the
kinematical group of noncommutative quantum mechanics,
satisfy a standard resolution of the identity (see (\ref{nc-resolid})). We shall also discuss
the relationship of these coherent states to the canonical coherent states (in this
case arising from the Weyl-Heisenberg group), for two degrees of freedom, and the fact
that these latter can be recovered from
the coherent states (\ref{coherentsttgalilei}) of noncommutative quantum mechanics in the
limit of $\vartheta = 0$.

\subsection{UIRs of the group $G_\text{\tiny Gal}^\text{\tiny ext}$}\label{uir-ext-gal}
The unitary irreducible representations of the extended Galilei group
$G_\text{\tiny Gal}^\text{\tiny ext}$ can be obtained from its projective unitary irreducible representations
worked out in \cite{bosejmpt}. We take, as mentioned earlier, both extension parameters
$m$ and $\lambda$ to be non-zero. The representation space is $L^2 (\mathbb R^2, d\bk)$
(momentum space representation).
Denoting the unitary representation operators by $\hat{U}_{m,\lambda}$, we have,
\begin{eqnarray}\label{unitaryirrerep}
\lefteqn{(\hat{U}_{m,\lambda}(\theta,\phi,R,b,\bv,\ba)
\hat{f})(\underbar{k})}\nonumber\\
&&=e^{i(\theta+\phi)}e^{i[\ba\cdot (\bk-\frac{1}{2}
m\bv )+\frac{b}{2m} \bk\cdot \bk
+\frac{\lambda}{2m}\bv\wedge\bk ]}s(R)\hat{f}
(R^{-1}(\bk-m\bv)),
\end{eqnarray}
for any $\quad \hat{f}\in L^{2}(\hat{\mathbb{R}}^{2},d\underbar{k})$. Here,
$s$ denotes the irreducible representation of the rotation
group in the rest frame (spin). It is useful to Fourier transform the above representation to
get its configuration space version (on $L^2 (\mathbb R^2, d\bx )$). A straightforward
computation, using Fourier transforms, leads to:
\begin{lem}\label{irrepconfig}
The unitary irreducible representations of  $G_\text{\tiny Gal}^\text{\tiny ext}$ in
the (two-dimensional) configuration space are given by
\begin{eqnarray}
\lefteqn{(U_{m,\lambda}(\theta,\phi,R,b,\bv,\ba )f)
(\bx)}\nonumber\\
&&=e^{i(\theta+\phi)}e^{i m(\bx+\frac{1}{2}\ba)\cdot\bv
}e^{-i\frac{b}{2m}\mathrm{\nabla}^{2}}s(R)f\left(R^{-1}
\left(\bx+\ba-\frac{\lambda}{2m}J\bv\right)\right),
\label{cord-rep}
\end{eqnarray}
where $\nabla^2 = \dfrac {\partial^2}{\partial x^2} + \dfrac {\partial^2}{\partial y^2}$, $J$ is the $2\times 2$ skew-symmetric matrix $J=\begin{pmatrix}0&-1\\1&0
\end{pmatrix}$ and $f\in L^{2}(\mathbb{R}^{2},d\bx)$.
\end{lem}

\subsection{Coherent states of the centrally extended (2+1)-Galilei group}\label{CSgal}
It is easy to see from (\ref{cord-rep}) that the representation $U_{m,\lambda}$
is {\em not square-integrable\/.} This means that there is no non-zero vector $\eta$ in the
representation space for which the function $f_\eta (g) =
\langle \eta \mid U_{m, \lambda}(g)\eta\rangle$ has finite $L^2$-norm, i.e., for all non-zero
$\eta \in L^2 (\mathbb R^2, d\bx )$,
$$ \int_{G_\text{\tiny Gal}^\text{\tiny ext}}\vert f_\eta (g)\vert^2 \; d\mu (g) =
   \infty \; , $$
$d\mu$ being the Haar measure.

On the other hand, the group composition law (\ref{group-comp-law}) reflects the fact that
the subgroup $H := \Theta\times\Phi\times SO(2)\times\mathcal{T}$, with generic group elements
$(\theta,\phi,R,b)$, is an abelian subgroup of $G_\text{\tiny Gal}^\text{\tiny ext}$.
The left coset space $X := G_\text{\tiny Gal}^\text{\tiny ext}/H$
is easily seen to be homeomorphic to  $\mathbb{R}^{4}$, corresponding to to the left
coset decomposition,
$$ (\theta , \phi , R, b, \bv , \ba ) = (0,0, \mathbb I_2, 0, \bv , \ba )
                (\theta , \phi , R, \mathbf 0 , \mathbf 0 )\; ,\quad (\mathbb I_2 = 2\times 2 \;
                \text{unit matrix}). $$
Writing $\bq$ for $\ba$ and replacing $\bv$ by $\bp := m\bv$, we  identify $X$ with the phase
space of the quantum system corresponding to the UIR $\hat{U}_{m,\lambda}$ and write its elements
as $(\bq,\bp)$. The homogeneous space carries an invariant measure under the natural action
of $G_\text{\tiny Gal}^\text{\tiny ext}$, which in these coordinates is just the Lebesgue
measure $d\bq\; d\bp$ on $\mathbb R^4$.
Also we define a section $\beta : X \longmapsto G_\text{\tiny Gal}^\text{\tiny ext}$,
\begin{equation}\label{sectionbundle}
\beta (\bq , \bp ) =(0,0,\mathbb I_2,0, \frac {\bp}m,\bq).
\end{equation}

We show next that the representation $U_{m, \lambda}$ is square-integrable mod$(\beta , H)$ in the
sense of \cite{coherentt} and hence construct coherent states on the homogeneous space (phase space)
$X$. Let $\chi \in L^2 (\mathbb R^2 , d\bx )$ be a fixed vector. At a later stage
(see Theorem \ref{CSquantizationofphasespacevar}) we shall need to impose a symmetry condition
on this vector, but at the moment we leave it arbitrary. For each phase space point
$(\bq , \bp)$ define the vector,
\begin{equation}
  \chi_{\bq , \bp} = U_{m, \lambda } (\beta (\bq , \bp ))\chi\; .
\label{cohst1}
\end{equation}
so that from (\ref{cord-rep}) and (\ref{sectionbundle}),
\begin{equation}\label{coherentsttgalilei}
\chi_{\bq,\bp}(\bx)=e^{i(\bx+\frac{1}{2}\bq )\;\cdot\bp}\chi\left(\bx+\bq -\frac{\lambda}{2m^{2}}
J\bp\right).
\end{equation}

\begin{lem}\label{resolutionofidentity}
For all $f, g \in L^2 (\mathbb R^2 )$, the vectors  $\chi_{\bq , \bp}$ satisfy
the square integrability condition
\begin{equation}
\int_{\mathbb{R}^{2}\times\mathbb{R}^{2}}\langle f \mid \chi_{\bq,\bp}
\rangle\langle\chi_{\bq ,\bp}\mid g \rangle \;d\bq\;d\bp = (2\pi)^{2}\Vert\chi\Vert^2
\langle f \mid g \rangle.
\label{nc-resolid}
\end{equation}
\end{lem}
The proof is given in the Appendix.
Additionally, in the course of the proof we have also established
 that the operator integral
$$  T = \int_{\mathbb R^2 \times \mathbb R^2} \vert \chi_{\bq , \bp}\rangle\langle
     \chi_{\bq , \bp} \vert \; d\bq\; d\bp \;  ,$$
 in (\ref{formal-op}) converges weakly to $T = 2\pi\;\Vert
\chi\Vert^2\; I$. Let us now define the vectors
\be
\eta = \frac 1{\sqrt{2\pi} \Vert\chi\Vert}\chi, \quad \text{and} \quad  \eta_{\bq , \bp} =
U(\beta (\bq , \bp ))\eta\; , \quad (\bq , \bp ) \in X\; .
\label{non-comm-CS}
\en
Then, as a consequence of the above lemma, we have proved the following theorem.
\begin{Theo}
The representation $U_{m, \lambda}$ in (\ref{cord-rep}), of the extended Galilei group
$G_\text{\tiny Gal}^\text{\tiny ext}$, is square integrable mod$\;(\beta , H)$ and the vectors
$\eta_{\bq , \bp}$ in (\ref{non-comm-CS}) form a set of coherent states defined on the
homogeneous space $X = G_\text{\tiny Gal}^\text{\tiny ext}/H$, satisfying the
resolution of the identity
\be
  \int_{\mathbb R^2 \times \mathbb R^2} \vert \eta_{\bq , \bp}\rangle\langle \eta_{\bq , \bp}
  \vert\;
     d\bq\; d\bp = I\; ,
\label{CSresolid}
\en
on $L^2 (\mathbb R^2 , d\bx )$.
\end{Theo}
Note that
\be
 \quad \eta_{\bq , \bp} (\bx) =
     e^{i(\bx+\frac{1}{2}\bq )\;\cdot\bp}\; \eta\left(\bx+\bq - \frac{\lambda}{2m^{2}}J\bp\right)\; .
\label{non-comm-CS2}
\en
We shall consider these coherent states to be the ones associated with non-commutative quantum
mechanics. Note that writing $\vartheta = \dfrac \lambda{m^2}$ as before, and letting $\vartheta
\rightarrow 0$, we recover the standard or {\em canonical coherent states} of ordinary quantum mechanics, if $\eta$ is chosen to be the gaussian wave function. Since this also corresponds to setting $\lambda = 0$, it is consistent with constructing the coherent states of the (2+1)-Galilei group with one central extension (using only the first of the two cocycles in (\ref{cocycles}), with mass parameter $m$).

  Let us emphasize again that the coherent states (\ref{non-comm-CS2}) are rooted in the
underlying symmetry group of noncommutative quantum mechanics and they are very different from the
ones introduced, for example, in \cite{scholtzt} and often used in the literature. These latter
coherent states are defined as   $\vert z ) = \vert z \rangle\langle z \vert, \; z \in \mathbb C$, where $\vert z \rangle$ is the usual canonical coherent state of ordinary quantum mechanics. If $\h$
denotes the Hilbert space of a one dimensional quantum system moving on the line, then $\vert z )$,  is an element of the space $\mathcal B_2 (\h)$ of Hilbert-Schmidt operators on $\h$ and this
 space is then taken to be the state space of noncommutative quantum mechanics. The coherent states $\vert z)$ satisfy a resolution of the identity which is also of a very different nature
from (\ref{CSresolid}). On $\mathcal B_2 (\h)$ the algebra of operators in (\ref{opsoscalgbr}) is realized by the operators $\widehat{\bQ}_i, \widehat{\bP}_i, \; i = 1,2$. These have the actions
\bea
\widehat{\bQ}_1 X = QX,  &\qquad &  \widehat{\bQ}_2 X = \vartheta PX, \nonumber\\
\widehat{\bP}_1 X = \hbar [P, X], & \qquad & \widehat{\bP}_2 X = -\frac {\hbar}\vartheta [Q, X] ,
\label{hilbschact}
\ena
on elements $X$ of $\mathcal B_2 (\h)$. The $Q$ and $P$ are two operators on
$\h$, satisfying the commutation relation $[Q, P] = iI_\h$. The state space with which we are working here and on which the operators (\ref{opsoscalgbr}) are realized, is $L^2 (\mathbb R^2, d\bx )$. It is not hard to see that the unitary {\em Wigner map}, $\mathcal W : \mathcal B_2 (\h) \longrightarrow
L^2 (\mathbb R^2, d\bx )$, given by
\be
  (\mathcal W X)(x,y) = \frac 1 {\sqrt{2\pi}}\; \text{Tr} [e^{-i(xQ + yP)}X]\; ,
\label{wigmap}
\en
transforms the set $\{ \widehat{\bQ}_i, \widehat{\bP}_i\}$ to the set  $\{ \tilde{Q}_i, \tilde{P}_i\}$ in (\ref{opsoscalgbr}). In other words, the formulation of noncommutative quantum mechanics on these two
state spaces are completely equivalent.

\subsection{Coherent state quantization on phase space leading to the noncommutative plane}
\label{csquantizationofgalgrp}
It has been already noted that we are identifying the homogeneous space
$X = G_\text{\tiny Gal}^\text{\tiny ext}/H$ with the phase space of the system. We shall now carry
out a coherent state quantization of functions on this phase space, using the above coherent states
of the extended Galilei group. It will turn out that such a quantization of the phase space variables
of position and momentum will lead precisely to the operators (\ref{opsoscalgbr}).

Recall that given a (sufficiently well behaved) function $f(\bq , \bp )$, its quantized version
$\hat{\mathcal{O}}_{f}$, obtained via coherent state quantization, is the operator (on
$L^2 (\mathbb R^2 , d\bx )$) given by the prescription (see, for example, \cite{aliengt} ),
\begin{equation}\label{CSquantscheme}
\hat{\mathcal{O}}_{f}=\int_{\mathbb{R}^{2}\times\mathbb{R}^{2}}f(\bq,\bp)
\vert\eta_{\bq,\bp}\rangle
\langle\eta_{\bq,\bp}\vert\; d\bq\;d\bp
\end{equation}
provided this operator is well-defined (again the integral being weakly defined).
The operators $\hat{\mathcal{O}}_{f}$ act on a
$g\in L^{2}(\mathbb{R}^{2},d\bx)$ in the following manner
\begin{equation}\label{actionofoperators}
(\hat{\mathcal{O}}_{f}g)(\bx)=\int_{\mathbb{R}^{2}\times\mathbb{R}^{2}}
f(\bq,\bp)\eta_{\bq,\bp}(\bx)\left[\int_{\mathbb{R}^{2}}
\overline{\eta_{\bq,\bp}(\bx^{\prime})}g(\bx^{\prime})
d\bx^{\prime}\right]\; d\bq\; d\bp\; .
\end{equation}
If we now take the function $f$ to be one of the coordinate functions, $f(\bq , \bp) = q_i, \; i = 1,2$,
or one of the momentum functions, $f(\bq , \bp ) =p_i, \; i = 1,2$, then the following theorem shows that
the resulting quantized operators
 $\hat{\mathcal{O}}_{q_i}$ and $\hat{\mathcal{O}}_{p_i}$
are exactly the ones given in (\ref{nc-comm-relns1}) for noncommutative quantum mechanics (with $\hbar = 1$) or the ones in (\ref{opsoscalgbr}), for the generators of the UIRs of $G_\text{\tiny Gal}^\text{\tiny ext}$ or of the noncommutative Weyl-Heisenberg group.

\begin{Theo}\label{CSquantizationofphasespacevar}
Let $\eta$ be a smooth function which satisfies the rotational invariance condition, $\eta (\bx ) = \eta (\Vert\bx \Vert )$, for all $\bx \in \mathbb R^2$. Then,
the operators  $\hat{\mathcal{O}}_{q_i}, \hat{\mathcal{O}}_{p_i} ,\;   i=1,2$, obtained by a
quantization of the phase space functions $q_{i}, p_{i}, \; i=1,2$, using the coherent states
(\ref{non-comm-CS}) of the (2+1)-centrally extended Galilei group, $G_\text{\tiny Gal}^\text{\tiny ext}$,
are given by
\bea
(\hat{\mathcal{O}}_{q_{1}}g)(\bx)=\left(x_{1}+\frac{i\lambda}{2 m^{2}}
\frac{\partial}{\partial x_{2}}\right)g(\bx) & \qquad &
(\hat{\mathcal{O}}_{q_{2}}g)(\bx) = \left(x_{2}-\frac{i\lambda}{2 m^{2}}
\frac{\partial}{\partial x_{1}}\right)g(\bx)\nonumber \\[2mm]
(\hat{\mathcal{O}}_{p_{1}}g)(\bx) = -i\frac{\partial}{\partial x_{1}}g(\bx)
& \qquad &
(\hat{\mathcal{O}}_{p_{2}}g)(\bx) = -i\frac{\partial}{\partial x_{2}}g(\bx),
\label{csquantops}
\ena
for $g \in L^{2}(\mathbb{R}^{2},d\bx)$, in the domain of these operators.
\end{Theo}

In (\ref{csquantops}) if we make the substitution $\vartheta = \dfrac \lambda{m^2}$, we get the
operators  (\ref{opsoscalgbr}) and the commutation relations of non-commutative quantum mechanics (with $\hbar = 1$):
\be \label{commtfornoncommplane}
[\hat{\mathcal{O}}_{q_{1}},\hat{\mathcal{O}}_{q_{2}}]=i\vartheta I ,\qquad
[\hat{\mathcal{O}}_{q_{i}},\hat{\mathcal{O}}_{p_{j}}]=i\delta_{ij}I ,\qquad
[\hat{\mathcal{O}}_{p_{i}},\hat{\mathcal{O}}_{p_{j}}]=0 .
\en
Moreover, it ought to be emphasized here that the rotational invariance of $\eta$, in the sense
that $\eta (\bx)  = \eta (\Vert \bx\Vert )$ was essential in deriving (\ref{csquantops}).

Two final remarks, before leaving this section, are in order.
First, the operators $Q_i , P_i$, appearing in (\ref{trnsfrmsecnd}),
together with the identity operator $I$, generate a representation of the Lie algebra of the Wey-Heisenberg group. Thus, it would seem that the operators  $\hat{Q}_i , \hat{P}_i$ are just a
different basis in this same algebra. However, this only appears to be so at the representation
level, in which both central elements of the extended Galilei group are mapped to the
identity operator. The two sets of operators, $Q_i , P_i$ and $\hat{Q}_i , \hat{P}_i$, in fact refer to the Lie algebras of two different
groups  namely, the $(2+1)$-Galilei groups with one and two extensions, respectively.
Moreover, the  set of commutation relations (\ref{nc-comm-relns1}), governing noncommutative quantum mechanics, is not unitary equivalent to that of standard quantum mechanics
(where $\vartheta = 0$). In the following section we  look at extensions of the Weyl-Heisenberg group which will throw more light on this issue.
As a second point, we note that the first commutation relation,
between $\hat{\mathcal{O}}_{q_{1}}$ and $\hat{\mathcal{O}}_{q_{2}}$ in
(\ref{commtfornoncommplane}) above, also implies that the two dimensional plane $\mathbb R^2$
becomes noncommutative as a result of quantization.

\section{Central extensions of the abelian group of translations in $\mathbb{R}^{4}$ and noncommutative quantum mechanics}\label{NCalgbrrevstd}
We start out with the abelian group of translations $G_{T}$ in $\mathbb{R}^{4}$, a generic element of which, denoted $({\bq},{\bp})$, obeys the group composition rule
\begin{equation}\label{grpoftrnsltn}
({\bq},{\bp})({\bq}^{\prime},{\bp}^{\prime})=({\bq}+{\bq}^{\prime},{\bp}+{\bp}^{\prime})\; .
\end{equation}
 At the  level of the Lie algebra, all the generators commute with each other. In order to arrive at quantum mechanics out of this abelian Lie group, and  to go further to obtain noncommutative quantum mechanics, we need to centrally extend this group of translations by some other abelian group, say by $\mathbb{R}$. In this section we will first discuss the double central extension of $G_{T}$  and  see that the double central extension by $\mathbb{R}$ yields the commutation relations (\ref{nc-comm-relns1}) of noncommutative quantum mechanics.  We will, next go a step further and extend $G_{T}$ triply by $\mathbb{R}$. The Lie algebra basis will be found to satisfy the
 additional commutation relation (\ref{non-comm-relns2}) between the momentum operators. We
 start by recalling some facts  about central extensions, following closely  the treatment of Bargmann in \cite{bargmannt}.

 Given a connected and simply connected Lie group $G$, the local exponents $\xi$ giving its central extensions are functions $\xi:G\times G\rightarrow \mathbb{R}$, obeying the following properties:
\begin{eqnarray}
\xi(g^{\prime\prime},g^{\prime})+\xi(g^{\prime\prime}g^{\prime},g)=\xi(g^{\prime\prime},g^{\prime}g)+\xi(g^{\prime},g)\label{defcentrlextfirst}\\
\xi(g,e)=0=\xi(e,g),\;\xi(g,g^{-1})=\xi(g^{-1},g)\label{defcentrlexsecond}.
\end{eqnarray}
We call the central extension trivial when the corresponding local exponent is simply a {\em coboundary} term,  in other words, when there exists a continuous function $\zeta:G\rightarrow \mathbb{R}$ such that
\begin{equation}\label{deftrivcentrlex}
\xi (g^{\prime},g) =  \xi_{cob} (g^{\prime},g) := \zeta(g^{\prime})+\zeta(g)-\zeta(g^{\prime}g).
\end{equation}
Two local exponents $\xi$ and $\xi^{\prime}$ are {\em equivalent} if they differ  by a coboundary term, i.e. $\xi^\prime (g^\prime,g)=\xi(g^\prime,g)+\xi_{cob}(g^\prime,g)$. A local exponent which is itself a  coboundary is said to be trivial and the corresponding extension of the group is called a trivial extension. Such an extension is isomorphic to the direct product group $\mathbb U(1) \times G$. Exponentiating the inequivalent local exponents yields the $\mathbb{U}(1)$ local factors or the familiar group multipliers, and the set of all such inequivalent multipliers form the well known second cohomology group $H^{2}(G,\mathbb{U}(1))$ of $G$.
\begin{Theo}\label{ineqvlocalexpgrptrns}
The three real valued functions $\xi$, $\xi^{\prime}$ and $\xi^{\prime\prime}$ on $G_{T}\times G_{T}$ given by
\begin{eqnarray}
\xi((q_{1},q_{2},p_{1},p_{2}),(q_{1}^{\prime},q_{2}^{\prime},p_{1}^{\prime},p_{2}^{\prime}))&=&\frac{1}{2}[q_{1}p_{1}^{\prime}+q_{2}p_{2}^{\prime}-p_{1}q_{1}^{\prime}-p_{2}q_{2}^{\prime}],\\
\xi^{\prime}((q_{1},q_{2},p_{1},p_{2}),(q_{1}^{\prime},q_{2}^{\prime},p_{1}^{\prime},p_{2}^{\prime}))&=&\frac{1}{2}[p_{1}p_{2}^{\prime}-p_{2}p_{1}^{\prime}],\\
\xi^{\prime\prime}((q_{1},q_{2},p_{1},p_{2}),(q_{1}^{\prime},q_{2}^{\prime},p_{1}^{\prime},p_{2}^{\prime}))&=&\frac{1}{2}[q_{1}q_{2}^{\prime}-q_{2}q_{1}^{\prime}],
\end{eqnarray}
are inequivalent local exponents for the group, $G_T$, of translations in $\mathbb{R}^{4}$ in the sense of (\ref{deftrivcentrlex}).
\end{Theo}

The proof is given in the Appendix.

\subsection{Double central extension of \texorpdfstring{$G_{T}$}{G_{T}}}\label{twocentrlext}
In this section, we study the doubly (centrally)  extended  group $\overline{\overline{G_{T}}}$ where the extension is achieved by means of the two multipliers $\xi$ and $\xi^{\prime}$ enumerated in Theorem \ref{ineqvlocalexpgrptrns}. The group composition rule for the extended group $\overline{\overline{G_{T}}}$ reads
\begin{eqnarray}\label{grplawdoublecentrlextns}
\lefteqn{(\theta,\phi,{\bq},{\bp})(\theta^{\prime},\phi^{\prime},{\bq}^{\prime},{\bp}^{\prime})}
   \nonumber\\
&&=(\theta+\theta^{\prime}+\frac{\alpha}{2}[\langle{\bq},{\bp}^{\prime}\rangle-
\langle{\bp},{\bq}^{\prime}\rangle],\phi+\phi^{\prime}+
\frac{\beta}{2}[{\bp}\wedge{\bp}^{\prime}],{\bq}+{\bq}^{\prime},{\bp}+{\bp}^{\prime}),
\end{eqnarray}
where $\bq = (q_1, q_2)$  and $\bp = (p_{1},p_{2})$. Also, $\langle.,.\rangle$ and $\wedge$ are defined as $\langle {\bq},{\bp}\rangle:=q_{1}p_{1}+q_{2}p_{2}$ and ${\bq}\wedge{\bp}:=q_{1}p_{2}-q_{2}p_{1}$ respectively.

A matrix representation for the group $\overline{\overline{G_{T}}}$ obeying the group multiplication rule (\ref{grplawdoublecentrlextns}) is given by the following $7\times 7$ upper triangular matrix
\begin{equation}\label{matrepdoublecentrlext}
(\theta,\phi,\bq,\bp)_{\alpha,\beta}=\begin{bmatrix}1&0&-\frac{\alpha}{2}p_{1}&-\frac{\alpha}{2}p_{2}&
\frac{\alpha}{2}q_{1}&\frac{\alpha}{2}q_{2}&\theta\\0&1&0&0&-\frac{\beta}{2}p_{2}
&\frac{\beta}{2}p_{1}&\phi\\0&0&1&0&0&0&q_{1}\\0&0&0&1&0&0&q_{2}\\0&0&0&0&1&0&p_{1}\\
0&0&0&0&0&1&p_{2}\\0&0&0&0&0&0&1\end{bmatrix}.
\end{equation}
Let us denote the generators of the Lie group $\overline{\overline{G_{T}}}$, or equivalently the basis of the associated Lie algebra, $\overline{\overline{\mathcal{G}_{T}}}$ by $\Theta, \Phi, Q_{1}, Q_{2}, P_{1}$ and $P_{2}$. These  generate the one-parameter subgroups  corresponding to the group parameters $\theta, \phi, p_{1}, p_{2}, q_{1}$ and $q_{2}$, respectively. The bilinear Lie brackets between the basis elements of $\overline{\overline{\mathcal{G}_{T}}}$ are given by
\begin{equation}\label{commutreldoubleext}
\begin{split}
&[P_{i},Q_{j}]=\alpha\delta_{i,j}\Theta,\;
[Q_{1},Q_{2}]=\beta\Phi,\quad
[P_{1},P_{2}]=0,\quad
[P_{i},\Theta]=0,\\
&[Q_{i},\Theta]=0,\quad
[P_{i},\Phi]=0,\quad
[Q_{i},\Phi]=0,\quad
[\Theta,\Phi]=0, \quad i,j =1,2\; .
\end{split}
\end{equation}
It is easily seen from (\ref{commutreldoubleext}) that $\Theta$ and $\Phi$ form the center of the algebra $\overline{\overline{\mathcal{G}_{T}}}$.
It is also noteworthy that, unlike in standard quantum mechanics, the two generators of space translation, $Q_1 , Q_2$, no longer commute, the noncommutativity of these two  generators being controlled by the central extension parameter $\beta$. It is in this context that it is reasonable to call the Lie group $\overline{\overline{G_{T}}}$ the {\em noncommutative Weyl-Heisenberg group} and the corresponding Lie algebra  the {\em noncommutative Weyl-Heisenberg algebra}.

We now proceed to find a unitary irreducible representation of  $\overline{\overline{G_{T}}}$. From the matrix representation (\ref{matrepdoublecentrlext}) we see that  $\overline{\overline{G_{T}}}$
is a nilpotent  Lie group. Hence, it is convenient to apply the orbit method of Kirillov (see \cite{Kirillovbook}) for finding the unitary dual of the group.

Switching to a slightly different notation, for computational convenience, we replace the group parameters $p_1, p_2, q_1, q_2, \theta$ and $\phi$ by $a_1, a_2, a_3, a_4, a_5$ and $a_6$, respectively. then a generic group element $g(a_1,a_2,a_3,a_4,a_5,a_6)$ is represented by the following matrix (compare with (\ref{matrepdoublecentrlext})):
\begin{equation}\label{matrepchangenotan}
g(a_1,a_2,a_3,a_4,a_5,a_6)=\begin{bmatrix}1&0&-\frac{\alpha}{2}a_{1}&-\frac{\alpha}{2}a_{2}&\frac{\alpha}{2}a_{3}&\frac{\alpha}{2}a_{4}&a_{5}\\0&1&0&0&-\frac{\beta}{2}a_{2}&\frac{\beta}{2}a_{1}&a_{6}\\0&0&1&0&0&0&x_{3}\\0&0&0&1&0&0&a_{4}\\0&0&0&0&1&0&a_{1}\\0&0&0&0&0&1&a_{2}\\0&0&0&0&0&0&1\end{bmatrix}.
\end{equation}
If $X_{1},X_{2},X_{3},X_{4},X_{5}$ and $X_{6}$ stand for the respective group generators,  a generic Lie algebra element can be written as $X=x^{1}X_{1}+x^{2}X_{2}+x^{3}X_{3}+x^{4}X_{4}+x^{5}X_{5}+x^{6}X_{6}$. In matrix notation, $X$ can be read off as
\begin{equation}\label{genrcalgbrelmntdoubleext}
X=\begin{bmatrix}0&0&-\frac{\alpha}{2}x^{1}&-\frac{\alpha}{2}x^{2}&\frac{\alpha}{2}x^{3}&\frac{\alpha}{2}x^{4}&x^{5}\\0&0&0&0&-\frac{\beta}{2}x^{2}&\frac{\beta}{2}x^{1}&x^{6}\\0&0&0&0&0&0&x^{3}\\0&0&0&0&0&0&x^{4}\\0&0&0&0&0&0&x^{1}\\0&0&0&0&0&0&x^{2}\\0&0&0&0&0&0&0\end{bmatrix}.
\end{equation}

An element $F\in\left(\overline{\overline{\mathcal{G}_{T}}}\right)^{*}$ with coordinates $\{X_{1},X_{2},X_{3},X_{4},X_{5},X_{6}\}$ is now represented by the following $7\times 7$ lower triangular matrix
\begin{equation}\label{dualalgbrelemntdoublext}
F=\begin{bmatrix}0&0&0&0&0&0&0\\0&0&0&0&0&0&0\\0&0&0&0&0&0&0\\0&0&0&0&0&0&0\\0&0&0&0&0&0&0\\0&0&0&0&0&0&0\\X_{5}&X_{6}&X_{3}&X_{4}&X_{1}&X_{2}&0\end{bmatrix},
\end{equation}
with the dual pairing being given as $\langle F,X\rangle=\hbox{tr}(FX)=\displaystyle\sum\limits_{i=1}^{6}x^{i}X_{i}$.
 Hence the coadjoint action $K$ of the underlying group $G_{T}$ on the dual Lie algebra $\left(\overline{\overline{\mathcal{G}_{T}}}\right)^{*}$ can be computed as
\begin{eqnarray}
\lefteqn{g(a_{1},a_{2},a_{3},a_{4},a_{5},a_{6})\;
F\; g(a_{1},a_{2},a_{3},a_{4},a_{5},a_{6})^{-1}}\nonumber\\
&&=\begin{bmatrix}*&*&*&*&*&*&*\\ *&*&*&*&*&*&*\\ *&*&*&*&*&*&*\\ *&*&*&*&*&*&*\\ *&*&*&*&*&*&*\\ *&*&*&*&*&*&*\\X_{5}^{\prime}&X_{6}^{\prime}&X_{3}^{\prime}&X_{4}^{\prime}&X_{1}^{\prime}&X_{2}^{\prime}&*\end{bmatrix},\label{nonzeroentrmatrx}
\end{eqnarray}
with
\begin{equation}\label{coadactiondoubleext}
\begin{split}
&X_{1}^{\prime}=X_{1}-\frac{\alpha}{2}a_{3}X_{5}+\frac{\beta}{2}a_{2}X_{6},\quad
 X_{2}^{\prime}=X_{2}-\frac{\alpha}{2}a_{4}X_{5}-\frac{\beta}{2}a_{1}X_{6},\\
&X_{3}^{\prime}=X_{3}+\frac{\alpha}{2}a_{1}X_{5},\quad
 X_{4}^{\prime}=X_{4}+\frac{\alpha}{2}a_{2}X_{5},\quad
 X_{5}^{\prime}=X_{5},\quad
 X_{6}^{\prime}=X_{6}.
\end{split}
\end{equation}

The required coadjoint action $K$ of the group on the dual algebra is therefore given by
\begin{eqnarray}\label{coadactnfinalfrmdouble}
\lefteqn{Kg(a_{1},a_{2},a_{3},a_{4},a_{5},a_{6})(X_{1},\; X_{2},X_{3},X_{4},X_{5},X_{6})}\nonumber\\
&&=(X_{1}-\frac{\alpha}{2}a_{3}X_{5}+\frac{\beta}{2}a_{2}X_{6},\; X_{2}-
\frac{\alpha}{2}a_{4}X_{5}-\frac{\beta}{2}a_{1}X_{6},\; X_{3}+\frac{\alpha}{2}a_{1}X_{5},\nonumber\\
&&\;\;\;\;X_{4}+\frac{\alpha}{2}a_{2}X_{5},\; X_{5},X_{6}).
\end{eqnarray}
The entries denoted by $*$'s in (\ref{nonzeroentrmatrx}) are some nonzero values that we are not interested in. From (\ref{coadactnfinalfrmdouble}) one observes that the two coordinates $X_{5}$ and $X_{6}$ remain unchanged under the coadjoint action. This is expected since they correspond to the center of the underlying algebra. The only two {\em polynomial invariants} in this case are just $P(F)=X_{5}$ and $Q(F)=X_{6}$. The coadjoint orbits are given by the set $S_{\rho,\sigma}$, for some
fixed real numbers $\rho , \sigma$, with
\begin{equation}\label{coadorbdef}
S_{\rho,\sigma}=\{F\in\left(\overline{\overline{\mathcal{G}_{T}}}\right)^{*}\mid P(F)=\rho,\; Q(F)=\sigma\}.
\end{equation}
Now, the first four coordinates of the vector on the right hand side of (\ref{coadactnfinalfrmdouble}) can be made zero by a suitable choice of the group parameters $a_{1},a_{2},a_{3},a_{4},a_{5}$ and $a_{6}$. Therefore, for nonzero values of $\rho$ and $\sigma$ in (\ref{coadorbdef}), the vector $(0,0,0,0,\rho,\sigma)$ will lie in a coadjoint orbit $S_{\rho,\sigma}$ of codimension 2. Since the dual algebra space is six dimensional, i.e., the coadjoint orbit in question is  4 dimensional and it passes through the point $(0,0,0,0,\rho,\sigma)$.

We next have to find the subalgebra, of correct dimension, subordinate to $F$ (see (\ref{dualalgbrelemntdoublext})). If we work with this appropiate polarizing subalgebra and solve the master equation (see \cite{Kirillovbook}), the representation we end up with will be irreducible
and unitary. The correct dimension of the polarizing subalgebra in this context turns out to be $\frac{2+6}{2}=4$. The maximal abelian subalgebra $\mathfrak{h}$ of the underlying algebra $\overline{\overline{\mathcal{G}_{T}}}$ serves as the appropiate poralizing subalgebra in this case, i.e. $\mathfrak{h}$ is the maximal subalgebra with $F|_{[\mathfrak{h},\mathfrak{h}]}=0$. A generic element of $\mathfrak{h}$ can be obtained from (\ref{genrcalgbrelmntdoubleext}) just by putting $x^{1}=x^{2}=0$ in there. A generic element of the corresponding abelian subgroup $H$ can be represented by the following matrix
\begin{equation}\label{subgrprepdoubleext}
h(\theta,\phi,\bq)=\begin{bmatrix}1&0&0&0&\frac{\alpha}{2}q_{1}&\frac{\alpha}{2}q_{2}&\theta\\0&1&0&0&0&0&\phi\\0&0&1&0&0&0&q_{1}\\0&0&0&1&0&0&q_{2}\\0&0&0&0&1&0&0\\0&0&0&0&0&1&0\\0&0&0&0&0&0&1\end{bmatrix}.
\end{equation}
We now choose a section $\delta:S=H\backslash\overline{\overline{G_{T}}} \rightarrow\overline{\overline{G_{T}}}$ with $\delta({\mathbf s})=\delta(s_{1},s_{2})$ being given by the following $7 \times 7$ matrix
\begin{equation}\label{sectionmatdoubleext}
\delta(s_{1},s_{2})=\begin{bmatrix}1&0&-\frac{\alpha}{2}s_{1}&-\frac{\alpha}{2}s_{2}&0&0&0\\0&1&0&0&-\frac{\beta}{2}s_{2}&\frac{\beta}{2}s_{1}&0\\0&0&1&0&0&0&0\\0&0&0&1&0&0&0\\0&0&0&0&1&0&s_{1}\\0&0&0&0&0&1&s_{2}\\0&0&0&0&0&0&1\end{bmatrix}.
\end{equation}
With all the relevant matrices at our disposal, we move on to solving the master equation, which in this case involves solving the matrix equation
\begin{eqnarray}
\lefteqn{\left(\begin{smallmatrix}1&0&-\frac{\alpha}{2}s_{1}&-\frac{\alpha}{2}s_{2}&0&0&0\\0&1&0&0&-\frac{\beta}{2}s_{2}&\frac{\beta}{2}s_{1}&0\\0&0&1&0&0&0&0\\0&0&0&1&0&0&0\\0&0&0&0&1&0&s_{1}\\0&0&0&0&0&1&s_{2}\\0&0&0&0&0&0&1\end{smallmatrix}\right)\left(\begin{smallmatrix}1&0&-\frac{\alpha}{2}p_{1}&-\frac{\alpha}{2}p_{2}&\frac{\alpha}{2}q_{1}&\frac{\alpha}{2}q_{2}&\theta\\0&1&0&0&-\frac{\beta}{2}p_{2}&\frac{\beta}{2}p_{1}&\phi\\0&0&1&0&0&0&q_{1}\\0&0&0&1&0&0&q_{2}\\0&0&0&0&1&0&p_{1}\\0&0&0&0&0&1&p_{2}\\0&0&0&0&0&0&1\end{smallmatrix}\right)}\nonumber\\
&&=\left(\begin{smallmatrix}1&0&-\frac{\alpha}{2}(p_{1}+s_{1})&-\frac{\alpha}{2}(p_{2}+s_{2})&\frac{\alpha}{2}q_{1}&\frac{\alpha}{2}q_{2}&\theta-\frac{\alpha}{2}q_{1}s_{1}-\frac{\alpha}{2}q_{2}s_{2}\\0&1&0&0&-\frac{\beta}{2}(p_{2}+s_{2})&\frac{\beta}{2}(p_{1}+s_{1})&\phi-\frac{\beta}{2}p_{1}s_{2}+\frac{\beta}{2}p_{2}s_{1}\\0&0&1&0&0&0&q_{1}\\0&0&0&1&0&0&q_{2}\\0&0&0&0&1&0&p_{1}+s_{1}\\0&0&0&0&0&1&p_{2}+s_{2}\\0&0&0&0&0&0&1\end{smallmatrix}\right)\label{doublefirst}\\
&&=\left(\begin{smallmatrix}1&0&0&0&\frac{\alpha}{2}A&\frac{\alpha}{2}B&C\\0&1&0&0&0&0&D\\0&0&1&0&0&0&A\\0&0&0&1&0&0&B\\0&0&0&0&1&0&0\\0&0&0&0&0&1&0\\0&0&0&0&0&0&1\end{smallmatrix}\right)\left(\begin{smallmatrix}1&0&-\frac{\alpha}{2}E&-\frac{\alpha}{2}F&0&0&0\\0&1&0&0&-\frac{\beta}{2}F&\frac{\beta}{2}E&0\\0&0&1&0&0&0&0\\0&0&0&1&0&0&0\\0&0&0&0&1&0&E\\0&0&0&0&0&1&F\\0&0&0&0&0&0&1\end{smallmatrix}\right)\nonumber\\
&&=\left(\begin{smallmatrix}1&0&-\frac{\alpha}{2}E&-\frac{\alpha}{2}F&\frac{\alpha}{2}A
&\frac{\alpha}{2}B&C+\frac{\alpha}{2}BF+\frac{\alpha}{2}AE\\0&1&0&0&-\frac{\beta}{2}F
&\frac{\beta}{2}E&D\\0&0&1&0&0&0&A\\0&0&0&1&0&0&B\\0&0&0&0&1&0&E\\0&0&0&0&0&1&F\\
0&0&0&0&0&0&1\end{smallmatrix}\right),\label{doublesecond}
\end{eqnarray}
for the unknowns $A,B,C,D,E$ and $F$. Comparing (\ref{doublefirst}) with (\ref{doublesecond}), one
gets
\begin{equation}\label{unknownsdoublexten}
\begin{split}
&A=q_{1},\quad
 B=q_{2},\quad
 E=p_{1}+s_{1},\quad
 F=p_{2}+s_{2},\\
&C=\theta-\alpha\langle\bq ,\bs+\frac{1}{2}\bp\rangle,\quad
 D=\phi-\frac{\beta}{2}\bp\wedge\bs.
\end{split}
\end{equation}
We recall that the coadjoint orbit vector, associated to which we found the polarizing algebra, was of the form $(0,0,0,0,\rho,\sigma)$. In view of (\ref{unknownsdoublexten}), we therefore have the following theorem
\begin{Theo}\label{uirofdoubleextend}
The noncommutative Weyl-Heisenberg group $\overline{\overline{G_{T}}}$ admits a unitary irreducible representation realized on $L^{2}(\mathbb{R}^{2}, d\bs)$ by the operators $U(\theta,\phi,\bq,\bp)$:
\begin{equation}\label{uirofdoublextns}
(U(\theta,\phi,\bq,\bp)f)(\bs)=\exp{i\left(\theta+\phi-\alpha\langle\bq,\bs+\frac{1}{2}
\bp\rangle-\frac{\beta}{2}\bp\wedge\bs\right)}f(\bs+\bp),
\end{equation}
where $f\in L^{2}(\mathbb{R}^{2},d\bs)$.
\end{Theo}
From the one-parameter unitary groups
$U(\theta,0,0,0,0,0), \;  U(0,0,q_{1},0,0,0), $ etc, we obtain the
their self-adjoint generators (on $L^{2}(\mathbb{R}^{2}, d\bs)$),  $\hat{\Theta}$, $\hat{\Phi}$, $\hat{P_{1}}$, $\hat{P_{2}}$, $\hat{Q_{1}}$ and $\hat{Q_{2}}$,  using the general formula
\begin{equation*}
\hat{X}_{\phi}=i{\frac{dU(\phi)}{d\phi}}\bigg{\vert}_{\phi=0}.
\end{equation*}
Thus, we have the following Hilbert space representation of the noncentral group generators
\begin{equation}\label{selfadjointrepsdoubleextn}
\begin{split}
&\hat{P}_{1}=\alpha s_{1},\quad
 \hat{Q}_{1}=\frac{\beta}{2}s_{2}+i\frac{\partial}{\partial s_{1}},\\
&\hat{P}_{2}=\alpha s_{2},\quad
 \hat{Q}_{2}=-\frac{\beta}{2}s_{1}+i\frac{\partial}{\partial s_{2}},
\end{split}
\end{equation}
while the two central generators $\hat{\Theta}$ and $\hat{\Phi}$ are both mapped to the Identity operator $\mathbb{I}_{\mathfrak{H}}$ of $\mathfrak{H}=L^{2}(\mathbb{R}^{2},d\bs)$.
An inverse Fourier transformation leads to the expressions, (on the coordinate Hilbert space
$L^2 (\mathbb R^2, d\bx )$)
\begin{equation}\label{Fourtransselfadrep}
\begin{split}
&\hat{P}_{1}=-i\alpha\frac{\partial}{\partial x},\quad
 \hat{P}_{2}=-i\alpha\frac{\partial}{\partial y},\\
&\hat{Q}_{1}=x -\frac{i\beta}{2}\frac{\partial}{\partial y},\quad
 \hat{Q}_{2}= y +\frac{i\beta}{2}\frac{\partial}{\partial x}.
\end{split}
\end{equation}
which coincide with (\ref{opsoscalgbr}) if we identify $\alpha$ with $\hbar$ and $-\beta$ with
$\vartheta$.

The commutation relations are now
\be\label{commuthilbertspcoperatdoublext}
[\hat{Q}_{i},\hat{P}_{j}]=i\alpha\delta_{i,j} \mathbb{I}_{\mathfrak{H}},\quad
[\hat{Q}_{1},\hat{Q}_{2}]=-i\beta\mathbb{I}_{\mathfrak{H}}, \quad
[\hat{P}_{1},\hat{P}_{2}]=0.
\en

If we now set $\alpha = \hbar$ and $-\beta = \vartheta$, we again retrieve the commutation
relations (\ref{nc-comm-relns1}) of noncommutative quantum mechanics. This means, that as in the
case of the Galilei group, an additional central extension of the Weyl-Heisenberg group leads to
non-commutative quantum mechanics.

\subsection{Triple central extension of \texorpdfstring{$G_{T}$}{G_{T}}}\label{threecentrlext}
In this section we study the triple central extension of $G_{T}$ by $\mathbb{R}$ and compute a unitary irreducible representation of the extended group $\overline{\overline{\overline{G_{T}}}}$. We will make use of all the three local exponents $\xi$, $\xi^{\prime}$ and $\xi^{\prime\prime}$ enumerated in Theorem \ref{ineqvlocalexpgrptrns} to do this triple extension. The group composition rule for the resulting triply extended Lie group $\overline{\overline{\overline{G_{T}}}}$ then reads
\begin{eqnarray}\label{grplawtriplyextendedgrp}
\lefteqn{(\theta,\phi,\psi,\bq,\bp)(\theta^{\prime},\phi^{\prime},\psi^{\prime},\bq^{\prime},\bp^{\prime})}\nonumber\\
&&=(\theta+\theta^{\prime}+\frac{\alpha}{2}[\langle\bq,\bp^{\prime}\rangle-\langle\bp,\bq^{\prime}\rangle],\phi+\phi^{\prime}+\frac{\beta}{2}[\bp\wedge\bp^{\prime}],\psi+\psi^{\prime}+\frac{\gamma}{2}[\bq\wedge\bq^{\prime}]\nonumber\\
&&\;\;\;,\bq+\bq^{\prime},\bp+\bp^{\prime}).
\end{eqnarray}
The matrix representation of $\overline{\overline{\overline{G_{T}}}}$, consistent with the above group law, is then  seen to be
\begin{equation}\label{matreptripleexten}
(\theta,\phi,\psi,\bq,\bp)_{\alpha,\beta,\gamma}=\begin{bmatrix}1&0&0&-\frac{\alpha}{2}p_{1}&-\frac{\alpha}{2}p_{2}&\frac{\alpha}{2}q_{1}&\frac{\alpha}{2}q_{2}&\theta\\0&1&0&0&0&-\frac{\beta}{2}p_{2}&\frac{\beta}{2}p_{1}&\phi\\0&0&1&-\frac{\gamma}{2}q_{2}&\frac{\gamma}{2}q_{1}&0&0&\psi\\0&0&0&1&0&0&0&q_{1}\\0&0&0&0&1&0&0&q_{2}\\0&0&0&0&0&1&0&p_{1}\\0&0&0&0&0&0&1&p_{2}\\0&0&0&0&0&0&0&1\end{bmatrix}.
\end{equation}
Let us denote the Lie algebra of $\overline{\overline{\overline{G_{T}}}}$ by $\overline{\overline{\overline{\mathcal{G}_{T}}}}$. Denoting the basis elements of $\overline{\overline{\overline{\mathcal{G}_{T}}}}$ by $\Theta,\Phi,\Psi,Q_{1},Q_{2},P_{1}$ and $P_{2}$, corresponding to the group parameters $\theta,\phi,\psi,p_{1},p_{2},q_{1}$ and $q_{2}$, respectively, we have the following Lie bracket relations between them
\begin{equation}\label{commutreltripleext}
\begin{split}
&[P_{i},Q_{j}]=\alpha\delta_{i,j}\Theta,\quad
 [Q_{1},Q_{2}]=\beta\Phi,\quad
 [P_{1},P_{2}]=\gamma\Psi,\quad
 [P_{i},\Theta]=0,\\
&[Q_{i},\Theta]=0,\quad
 [P_{i},\Phi]=0,\quad
 [Q_{i},\Phi]=0,\quad
 [P_{i},\Psi]=0,\\
&[Q_{i},\Psi]=0,\quad
 [\Theta,\Phi]=0,\quad
 [\Phi,\Psi]=0,\quad
 [\Theta,\Psi]=0, \quad i,j =1,2\; .
\end{split}
\end{equation}
In addition to the two central elements $\Theta$ and $\Phi$ appearing in  the double extension case (see (\ref{commutreldoubleext})), we have a third central element $\Psi$ in (\ref{commutreltripleext}), which makes the two generators $P_{1}$ and $P_{2}$ noncommutative as well, with the noncommutativity being controlled by the extension parameter $\gamma$. We shall call this centrally extended Lie group $\overline{\overline{\overline{G_{T}}}}$ the {\em triply extended group of translations} and the corresponding Lie algebra $\overline{\overline{\overline{\mathcal{G}_{T}}}}$ the {\em triply extended algebra of translations}.

It remains now to find a unitary irreducible representation of the group $\overline{\overline{\overline{G_{T}}}}$. In doing so we will be following exactly the same course as for the UIR of the  $\overline{\overline{G_{T}}}$ in Section \ref{twocentrlext}. Since $\overline{\overline{\overline{G_{T}}}}$ is also a nilpotent Lie group, (see (\ref{matreptripleexten})), we again apply the orbit method of Kirillov.

 We again change  notations  and replace the group parameters $p_{1},p_{2},q_{1}$, $q_{2},\theta,\phi$ and $\psi$ by $a_{1},a_{2},a_{3},a_{4},a_{5},a_{6}$ and $a_{7}$, respectively. Then, a generic group element  has the matrix representation
\begin{equation}\label{changednotdoubleexten}
g(a_{1},a_{2},a_{3},a_{4},a_{5},a_{6},a_{7})=\begin{bmatrix}1&0&0&-\frac{\alpha}{2}a_{1}&-\frac{\alpha}{2}a_{2}&\frac{\alpha}{2}a_{3}&\frac{\alpha}{2}a_{4}&a_{5}\\0&1&0&0&0&-\frac{\beta}{2}a_{2}&\frac{\beta}{2}a_{1}&a_{6}\\0&0&1&-\frac{\gamma}{2}a_{4}&\frac{\gamma}{2}a_{3}&0&0&a_{7}\\0&0&0&1&0&0&0&a_{3}\\0&0&0&0&1&0&0&a_{4}\\0&0&0&0&0&1&0&a_{1}\\0&0&0&0&0&0&1&a_{2}\\0&0&0&0&0&0&0&1\end{bmatrix}.
\end{equation}
 Denoting by $X_{1},X_{2},X_{3},X_{4},X_{5},X_{6}$ and $X_{7}$, the respective group generators, and writing a a generic Lie algebra element as $X=x^{1}X_{1}+x^{2}X_{2}+x^{3}X_{3}+x^{4}X_{4}+x^{5}X_{5}+x^{6}X_{6}+x^{7}X_{7}$, we have the matrix
\begin{equation}\label{liealgbrelemnttripleext}
X=\begin{bmatrix}0&0&0&-\frac{\alpha}{2}x^{1}&-\frac{\alpha}{2}x^{2}&\frac{\alpha}{2}x^{3}&\frac{\alpha}{2}x^{4}&x^{5}\\0&0&0&0&0&-\frac{\beta}{2}x^{2}&\frac{\beta}{2}x^{1}&x^{6}\\0&0&0&-\frac{\gamma}{2}x^{4}&\frac{\gamma}{2}x^{3}&0&0&x^{7}\\0&0&0&0&0&0&0&x^{3}\\0&0&0&0&0&0&0&x^{4}\\0&0&0&0&0&0&0&x^{1}\\0&0&0&0&0&0&0&x^{2}\\0&0&0&0&0&0&0&0\end{bmatrix}.
\end{equation}
We represent an element $F\in\left(\overline{\overline{\overline{\mathcal{G}_{T}}}}\right)^{*}$ with the following $8\times 8$ lower tringular matrix
\begin{equation}\label{dualalgbrelemnttripleextnsion}
F=\begin{bmatrix}0&0&0&0&0&0&0&0\\0&0&0&0&0&0&0&0\\0&0&0&0&0&0&0&0\\-\frac{2}{\alpha}X_{1}&0&0&0&0&0&0&0\\-\frac{2}{\alpha}X_{2}&0&0&0&0&0&0&0\\0&0&0&0&0&0&0&0\\0&0&0&0&0&0&0&0\\X_{5}&X_{6}&X_{7}&X_{3}&X_{4}&0&0&0\end{bmatrix},
\end{equation}
where the dual pairing is given by $\langle F,X\rangle=\hbox{tr}(FX)=\displaystyle\sum\limits_{i=1}^{7}x^{i}X_{i}$. Therefore, the coadjoint action of the underlying Lie group $\overline{\overline{\overline{G_{T}}}}$ on the corresponding dual Lie algebra $\left(\overline{\overline{\overline{\mathcal{G}_{T}}}}\right)^{*}$ follows from the following computation
\begin{eqnarray}
\lefteqn{g(a_{1},a_{2},a_{3},a_{4},a_{5},a_{6},a_{7})Fg(a_{1},a_{2},a_{3},a_{4},a_{5},a_{6},a_{7})^{-1}}\nonumber\\
&&=\begin{bmatrix}*&*&*&*&*&*&*&*\\ *&*&*&*&*&*&*&*\\ *&*&*&*&*&*&*&*\\-\frac{2}{\alpha}X_{1}^{\prime}&*&*&*&*&*&*&*\\-\frac{2}{\alpha}X_{2}^{\prime}&*&*&*&*&*&*&*\\ *&*&*&*&*&*&*&*\\ *&*&*&*&*&*&*&*\\X_{5}^{\prime}&X_{6}^{\prime}&X_{7}^{\prime}&X_{3}^{\prime}&X_{4}^{\prime}&*&*&*\end{bmatrix},\label{coadactionfordoubleexten}
\end{eqnarray}
with
\begin{equation}\label{coadactionreldoubleext}
\begin{split}
&X_{1}^{\prime}=-\frac{2}{\alpha}X_{1}+a_{3}X_{5},\;\;
 X_{2}^{\prime}=-\frac{2}{\alpha}X_{2}+a_{4}X_{5},\;\;
 X_{3}^{\prime}=\frac{\alpha}{2}a_{1}X_{5}+\frac{\gamma}{2}a_{4}X_{7}+X_{3},\\
&X_{4}^{\prime}=\frac{\alpha}{2}a_{2}X_{5}-\frac{\gamma}{2}a_{3}X_{7}+X_{4},\;\;
 X_{5}^{\prime}=X_{5},\;\;
 X_{6}^{\prime}=X_{6},\;\;
 X_{7}^{\prime}=X_{7}.
\end{split}
\end{equation}
The required coadjoint action of the group on the dual algebra is therefore given by
\begin{eqnarray}\label{coadacttripleextens}
\lefteqn{Kg(a_{1},a_{2},a_{3},a_{4},a_{5},a_{6},a_{7})(X_{1},X_{2},X_{3},X_{4},X_{5},X_{6},X_{7})}\nonumber\\
&&=(-\frac{2}{\alpha}X_{1}+a_{3}X_{5},-\frac{2}{\alpha}X_{2}+a_{4}X_{5},\frac{\alpha}{2}a_{1}X_{5}+\frac{\gamma}{2}a_{4}X_{7}+X_{3}\nonumber\\
&&\;\;\;\;,\frac{\alpha}{2}a_{2}X_{5}-\frac{\gamma}{2}a_{3}X_{7}+X_{4},X_{5},X_{6},X_{7}).
\end{eqnarray}
The nonzero entries denoted by $*$'s in (\ref{coadactionfordoubleexten}) are of no interest to us. From (\ref{coadacttripleextens}), one observes that the three dual algebra coordinates $X_{5},X_{6}$ and $X_{7}$ remain unaltered under the coadjoint action of the underlying group element, coming as they do from the center of the Lie algebra. We therefore have three polynomial invariants in our theory given by $P(F)=X_{5}$, $Q(F)=X_{6}$ and $R(F)=X_{7}$. The coadjoint orbits in this case are given by the sets $S_{\rho,\sigma,\tau}$ with
\begin{equation}\label{coadorbstripleexten}
S_{\rho,\sigma,\tau}=\{F\in\left(\overline{\overline{\overline{\mathcal{G}_{T}}}}\right)^{*}\mid P(F)=\rho,\; Q(F)=\sigma,\; R(F)=\tau\}.
\end{equation}
It is also obvious from (\ref{coadacttripleextens}) that by choosing $a_{1}$, $a_{2}$, $a_{3}$ and $a_{4}$ in a suitable manner, we can make all of $X_{1}^{\prime}$, $X_{2}^{\prime}$, $X_{3}^{\prime}$ and $X_{4}^{\prime}$ vanishing at the same time. Therefore, for nonzero values of $\rho$, $\sigma$ and $\tau$, the vector $(0,0,0,0,\rho,\sigma,\tau)$ will always lie in the coadjoint orbit $S_{\rho,\sigma,\tau}$ of codimension 3. In other words, the underlying coadjoint orbit $S_{\rho,\sigma,\tau}$ turns out to be  $4$ dimensional which passes through the point $(0,0,0,0,\rho,\sigma,\tau)$ of the dual algebra space.

We now have to find the maximal subalgebra subordinate to $F$ given by (\ref{dualalgbrelemnttripleextnsion}). This maximal subalgebra or the polarizing subalgebra turns out to be of the correct dimension $\frac{3+7}{2}=5$ and hence, the representation for $\overline{\overline{\overline{G_{T}}}}$ that we end up with using the orbit method will be irreducible and unitary. As in the case of $\overline{\overline{G_{T}}}$, the maximal abelian subalgebra $\mathfrak{h}$ of the Lie algebra  $\overline{\overline{\overline{\mathcal{G}_{T}}}}$ serves as the polarizing subalgebra. A generic element of the corresponding abelian subgroup $H$ can be represented by the following $8\times 8$ matrix
\begin{equation}\label{abeliansubgrptripleext}
h(\theta,\phi,\psi,p_{1},q_{2})=\begin{bmatrix}1&0&0&-\frac{\alpha}{2}p_{1}&0&0&\frac{\alpha}{2}q_{2}&\theta\\0&1&0&0&0&0&\frac{\beta}{2}p_{1}&\phi\\0&0&1&-\frac{\gamma}{2}q_{2}&0&0&0&\psi\\0&0&0&1&0&0&0&0\\0&0&0&0&1&0&0&q_{2}\\0&0&0&0&0&1&0&p_{1}\\0&0&0&0&0&0&1&0\\0&0&0&0&0&0&0&1\end{bmatrix}.
\end{equation}
Then the section $\delta:H\backslash\overline{\overline{\overline{G_{T}}}}\rightarrow\overline{\overline{\overline{G_{T}}}}$ will be represented by the following matrix
\begin{equation}\label{sectiontripleextens}
\delta(r_{1},s_{2})=\begin{bmatrix}1&0&0&0&-\frac{\alpha}{2}s_{2}&\frac{\alpha}{2}r_{1}&0&0\\0&1&0&0&0&-\frac{\beta}{2}s_{2}&0&0\\0&0&1&0&\frac{\gamma}{2}r_{1}&0&0&0\\0&0&0&1&0&0&0&r_{1}\\0&0&0&0&1&0&0&0\\0&0&0&0&0&1&0&0\\0&0&0&0&0&0&1&s_{2}\\0&0&0&0&0&0&0&1\end{bmatrix}.
\end{equation}
Thus, we again have to solve the master equation,
\begin{eqnarray}\label{solnofmastreqntripleextns}
\lefteqn{\left(\begin{smallmatrix}1&0&0&0&-\frac{\alpha}{2}s_{2}&\frac{\alpha}{2}r_{1}&0&0\\0&1&0&0&0&-\frac{\beta}{2}s_{2}&0&0\\0&0&1&0&\frac{\gamma}{2}r_{1}&0&0&0\\0&0&0&1&0&0&0&r_{1}\\0&0&0&0&1&0&0&0\\0&0&0&0&0&1&0&0\\0&0&0&0&0&0&1&s_{2}\\0&0&0&0&0&0&0&1\end{smallmatrix}\right)\left(\begin{smallmatrix}1&0&0&-\frac{\alpha}{2}p_{1}&-\frac{\alpha}{2}p_{2}&\frac{\alpha}{2}q_{1}&\frac{\alpha}{2}q_{2}&\theta\\0&1&0&0&0&-\frac{\beta}{2}p_{2}&\frac{\beta}{2}p_{1}&\phi\\0&0&1&-\frac{\gamma}{2}q_{2}&\frac{\gamma}{2}q_{1}&0&0&\psi\\0&0&0&1&0&0&0&q_{1}\\0&0&0&0&1&0&0&q_{2}\\0&0&0&0&0&1&0&p_{1}\\0&0&0&0&0&0&1&p_{2}\\0&0&0&0&0&0&0&1\end{smallmatrix}\right)}\nonumber\\
&&=\left(\begin{smallmatrix}1&0&0&-\frac{\alpha}{2}p_{1}&-\frac{\alpha}{2}(p_{2}+s_{2})&\frac{\alpha}{2}(q_{1}+r_{1})&\frac{\alpha}{2}q_{2}&\theta-\frac{\alpha}{2}q_{2}s_{2}+\frac{\alpha}{2}p_{1}r_{1}\\0&1&0&0&0&-\frac{\beta}{2}(p_{2}+s_{2})&\frac{\beta}{2}p_{1}&\phi-\frac{\beta}{2}p_{1}s_{2}\\0&0&1&-\frac{\gamma}{2}q_{2}&\frac{\gamma}{2}(q_{1}+r_{1})&0&0&\psi+\frac{\gamma}{2}q_{2}r_{1}\\0&0&0&1&0&0&0&q_{1}+r_{1}\\0&0&0&0&1&0&0&q_{2}\\0&0&0&0&0&1&0&p_{1}\\0&0&0&0&0&0&1&p_{2}+s_{2}\\0&0&0&0&0&0&0&1\end{smallmatrix}\right)\label{mastereqnfirsttripleext}\\
&&=\left(\begin{smallmatrix}1&0&0&-\frac{\alpha}{2}A&0&0&\frac{\alpha}{2}B&C\\0&1&0&0&0&0&\frac{\beta}{2}A&D\\0&0&1&-\frac{\gamma}{2}B&0&0&0&E\\0&0&0&1&0&0&0&0\\0&0&0&0&1&0&0&B\\0&0&0&0&0&1&0&A\\0&0&0&0&0&0&1&0\\0&0&0&0&0&0&0&1\end{smallmatrix}\right)\left(\begin{smallmatrix}1&0&0&0&-\frac{\alpha}{2}F&\frac{\alpha}{2}G&0&0\\0&1&0&0&0&-\frac{\beta}{2}F&0&0\\0&0&1&0&\frac{\gamma}{2}G&0&0&0\\0&0&0&1&0&0&0&G\\0&0&0&0&1&0&0&0\\0&0&0&0&0&1&0&0\\0&0&0&0&0&0&1&F\\0&0&0&0&0&0&0&1\end{smallmatrix}\right)\nonumber\\
&&=\left(\begin{smallmatrix}1&0&0&-\frac{\alpha}{2}A&-\frac{\alpha}{2}F&\frac{\alpha}{2}G&\frac{\alpha}{2}B&C+\frac{\alpha}{2}BF-\frac{\alpha}{2}GA\\0&1&0&0&0&-\frac{\beta}{2}F&\frac{\beta}{2}A&D+\frac{\beta}{2}AF\\0&0&1&-\frac{\gamma}{2}B&\frac{\gamma}{2}G&0&0&E-\frac{\gamma}{2}BG\\0&0&0&1&0&0&0&G\\0&0&0&0&1&0&0&B\\0&0&0&0&0&1&0&A\\0&0&0&0&0&0&1&F\\0&0&0&0&0&0&0&1\end{smallmatrix}\right).\label{mastereqnsecondtripleext}
\end{eqnarray}
The unknowns $A,B,C,D,E,F$ and $G$ can easily be computed by comparing (\ref{mastereqnfirsttripleext}) with (\ref{mastereqnsecondtripleext}). We get
\begin{equation}\label{unknownstrpleextens}
\begin{split}
&A=p_{1},\;
 B=q_{2},\;
 G=r_{1}+q_{1},\;
 F=s_{2}+p_{2},\\
&C=\theta-\alpha q_{2}s_{2}+\alpha p_{1}r_{1}+\frac{\alpha}{2}q_{1}p_{1}-\frac{\alpha}{2}q_{2}p_{2},\\
&D=\phi-\beta p_{1}s_{2}-\frac{\beta}{2}p_{1}p_{2},\;
 E=\psi+\gamma q_{2}r_{1}+\frac{\gamma}{2}q_{1}q_{2}.
\end{split}
\end{equation}
Now, the dual algebra vector lying in the underlying four dimensional coadjoint orbit was found to be $(0,0,0,0,\rho,\sigma,\tau)$. In light of (\ref{unknownstrpleextens}), we therefore have the following theorem
\begin{Theo}\label{uiroftripleextend}
The triply extended group of translations $\overline{\overline{\overline{G_{T}}}}$ admits a unitary irreducible representation realized on $L^{2}(\mathbb{R}^{2})$. The explicit form of the representation is given by
\begin{eqnarray}\label{uiroftriplextns}
\lefteqn{(U(\theta,\phi,\psi,q_{1},q_{2},p_{1},p_{2})f)(r_{1},s_{2})}\nonumber\\
&&=e^{i(\theta-\alpha q_{2}s_{2}+\alpha p_{1}r_{1}+\frac{\alpha}{2}q_{1}p_{1}-\frac{\alpha}{2}q_{2}p_{2})}e^{i(\phi-\beta p_{1}s_{2}-\frac{\beta}{2}p_{1}p_{2})}\nonumber\\
&&\times e^{i(\psi+\gamma q_{2}r_{1}+\frac{\gamma}{2}q_{2}q_{1})}f(r_{1}+q_{1},s_{2}+p_{2}),
\end{eqnarray}
where $f\in L^{2}(\mathbb{R}^{2},dr_{1}ds_{2})$.
\end{Theo}

Now, let us take the Fourier transform of (\ref{uiroftriplextns}) with respect to the first coordinate $r_{1}$ and call the transformed coordinate $s_{1}$. The noncentral generators of $\overline{\overline{\overline{G_{T}}}}$ can be represented by self adjoint operators defined on $L^{2}(\mathbb{R}^{2},d\bs)$ in the following manner
\begin{equation}\label{selfadreptrpleext}
\begin{split}
&\hat{P}_{1}=-s_{1},\quad
 \hat{Q}_{1}=\beta s_{2}-i\alpha \frac{\partial}{\partial s_{1}},\\
&\hat{P}_{2}=\alpha s_{2}-i\gamma\frac{\partial}{\partial s_{1}},\quad
 \hat{Q}_{2}=i\frac{\partial}{\partial s_{2}},
 \end{split}
\end{equation}
while the three central elements $\Theta,\Phi$ and $\Psi$ of the corresponding Lie algebra $\overline{\overline{\overline{\mathcal{G}_{T}}}}$ are all mapped to the identity operator $\mathbb{I}_{\mathfrak{H}}$ of the uderlying Hilbert space $\mathfrak{H}=L^{2}(\mathbb{R}^{2},d\bs)$.
The corresponding commutation relations now read
\be\label{commureltripleext}
[\hat{Q}_{i},\hat{P}_{j}]=i\alpha\delta_{i,j}\mathbb{I}_{\mathfrak{H}},\quad
[\hat{Q}_{1},\hat{Q}_{2}]=-i\beta\mathbb{I}_{\mathfrak{H}},\quad
[\hat{P}_{1},\hat{P}_{2}]=-i\gamma\mathbb{I}_{\mathfrak{H}}.
\en

 Once again, if we write $\alpha = \hbar,\;  -\beta = \vartheta$ and replace $\gamma$ by $-\gamma$ we
 recover (\ref{nc-comm-relns1}) together with (\ref{non-comm-relns2}), the additional central extension making the two momentum operators noncommuting.

\section{Conclusion and outlook}
We have derived the commutation relations between the position and momentum operators of noncommuttive quantum mechanics by three different means: using the appropriate unitary irreducible representations of the centrally extended (2+1)-Galilei group $G_\text{\tiny Gal}^\text{\tiny ext}$, of
the doubly extended group $\overline{\overline{G_{T}}}$,  of translations
of $\mathbb R^4$, and by a coherent state quantization of the classical phase space variables of
position and momentum, using the coherent states of $G_\text{\tiny Gal}^\text{\tiny ext}$. It is
not hard to see, from the expressions for the unitary representations (\ref{unitaryirrerep}) and
(\ref{uirofdoublextns}), that the same commutation relations could also be obtained by a coherent state quantization, using the coherent states of $\overline{\overline{G_{T}}}$ (which could
be similarly constructed). There is, as usual a positive operator valued (POV) measure
naturally associated to the coherent states (\ref{non-comm-CS}). Indeed, for any measurable set
$\Delta$ of $\mathbb R^2$ (phase space), we can associate the positive operator
$$
   a(\Delta ) = \int_{\Delta} \vert\eta_{\bq, \bp}\rangle\langle \eta_{\bq , \bp} \vert\;
      d\bq\; d\bp \; . $$
These define localization operators on phase space, whose marginals in $\bq$ and $\bp$
should then give
localization operators in configuration and momentum spaces, respectively. For the canonical coherent
states and standard quantum mechanics, such operators have been studied extensively,
in e.g., \cite{aliqmpspt,busgrablaht}. There, one understands these localization
operators in an extended or unsharp sense. It would be interesting to do a similar study for the present case.

\section{Appendix}\label{sec-app}
In this Appendix we collect the proofs of some of the results quoted in the paper.

\prf {\bf of Lemma \ref{resolutionofidentity}.}

We start out by taking two compactly supported  and infinitely differentiable
functions  $f, g\in L^{2}(\mathbb{R}^{2},d\bx)$.
Then,
\begin{eqnarray}\label{resident}
\lefteqn{\int_{\mathbb{R}^{2}\times\mathbb{R}^{2}}\langle f\mid \chi_{\bq,\bp}\rangle
\langle\chi_{\bq,\bp}\mid g\rangle \;d\bq\;d\bp}\nonumber\\[2mm]
&&=\int_{\mathbb R^2\times\mathbb R^2}d\bq\;d\bp\;
\bigg[\int_{\mathbb R^2\times\mathbb R^2}e^{i(\bx-
\bx^\prime )\cdot \bp }\chi\left(\bx +
\bq   - \frac{\lambda}{2m^{2}}J\bp\right)
\overline{\chi\left(\bx^{\prime}+\bq -
\frac{\lambda}{2m^{2}}J\bp\right)}\nonumber\\[2mm]
&&\times \overline{f(\bx)}g(\bx^{\prime})\;
d\bx\;d\bx^{\prime}\bigg]
\end{eqnarray}
Making the change of variables, $\bq-\dfrac{\lambda}{2m^{2}}J\bp = \bq^\prime$,
\begin{eqnarray}
\lefteqn{\int_{\mathbb{R}^{2}\times\mathbb{R}^{2}}\langle f \mid \chi_{\bq,
\bp}\rangle\langle\chi_{\bq,\bp}\mid g\rangle\;
d\bq\;d\bp}\nonumber\\[2mm]
&&=\int_{\mathbb{R}^{2}\times\mathbb{R}^{2}}d\bq^{\prime}\;d\bp\;
\left[\int_{\mathbb{R}^{2}\times\mathbb{R}^{2}}e^{i(\bx-
\bx^{\prime})\cdot\bp}\chi(\bx +\bq^{\prime})
\overline{\chi(\bx^{\prime}+\bq^{\prime})}
g(\bx^{\prime})\overline{f(\bx)}\;
d\bx\;d\bx^{\prime}\right]\nonumber\\[2mm]
&&=(2\pi)^{2}\int_{\mathbb{R}^{2}}d\bq^{\prime}\left[\int_{\mathbb{R}^{2}
\times\mathbb{R}^{2}}\delta(\bx -\bx^{\prime})\chi(\bx +
\bq^{\prime})\overline{\chi(\bx^{\prime}+\bq^{\prime})}
g(\bx^{\prime})\overline{f(\bx)}\;d\bx\;
d\bx^{\prime}\right]\nonumber\\[2mm]
&&=(2\pi)^{2}\int_{\mathbb{R}^{2}}d\bq^{\prime}\;\left[\int_{\mathbb{R}^{2}}
\chi(\bx +\bq^{\prime})\overline{\chi(\bx +
\bq^{\prime})}g(\bx)\overline{f(\bx)}\;d\bx\right]\nonumber\\[2mm]
&&=(2\pi)^{2}\|\chi\|^{2}\langle f \mid g\rangle,
\end{eqnarray}
the change in the order of integration and the introduction of the delta measure
being easily justified in view of the compact supports and smoothness property of
the functions $f$ and $g$. Thus, introducing the formal operator
\be
  T = \int_{\mathbb R^2 \times \mathbb R^2} \vert \chi_{\bq , \bp}\rangle\langle
     \chi_{\bq , \bp} \vert \; d\bq\; d\bp \;  ,
\label{formal-op}
\en
we see that for functions $f,g$ of the chosen type,
$$ \langle f \vert T g\rangle = 2\pi \Vert \chi\Vert^2\; \langle f \vert I g\rangle\;, $$
$I$ being the identity operator on $L^2 (\mathbb R^2 , d\bx )$. But since the compactly supported
and infinitely differentiable functions are dense in $L^2 (\mathbb R^2, d\bx )$, we use the
continuity of the scalar product to extend the above equality to arbitrary pairs of
functions $f, g$ in $L^2 (\mathbb R^2, d\bx )$, thus proving the lemma.
\qed

\bigskip
\prf {\bf of Theorem \ref{CSquantizationofphasespacevar}.}

We only work out the derivation of the first of the above equations, the others being
obtained in similar ways.
By (\ref{non-comm-CS2}) and  (\ref{actionofoperators})
\begin{eqnarray}\label{proofthmfirst}
\lefteqn{(\hat{\mathcal{O}}_{q_{1}}g)( \bx )}\nonumber\\
&&=\int_{\mathbb{R}^{2}\times\mathbb{R}^{2}}q_{1} \;  \bigg[\int_{\mathbb{R}^{2}}e^{i (\bx-
\bx^{\prime})\cdot \bp }\;\eta\;\left(\bx+\bq-
\frac{\lambda}{2 m^{2}}J\bp\right)\nonumber\\[2mm]
&&\times\overline{\eta\left(\bx^{\prime}+\bq -
\frac{\lambda}{2m^{2}}J\bp\right)}g(\bx^{\prime})\;d\bx^{\prime}\bigg]\;d\bq\;d\bp\;.
\end{eqnarray}
Making the change of variables $\bq -\dfrac{\lambda}{2m^{2}}J\bp =
\bq^\prime$, and noting the form of the skew-symmetric matrix  $J$ from
Lemma (\ref{irrepconfig}), we have
$$
q_{1}^{\prime} = q_{1}  + \frac{\lambda}{2m^{2}}p_{2},\qquad
q_{2}^{\prime} =  q_{2} - \frac{\lambda}{2m^{2}}p_{1}\; ,$$
using which (\ref{proofthmfirst}) becomes
\begin{eqnarray}\label{proofthmseond}
\lefteqn{(\hat{\mathcal{O}}_{q_{1}}g)(\bx)}\nonumber\\
&&=\int_{\mathbb{R}^{2}\times\mathbb{R}^{2}}\!\!\left(q_{1}^{\prime}-
\frac{\lambda}{2m^{2}}p_{2}\;\right)\left[\int_{\mathbb{R}^{2}}
e^{i(\bx - \bx^{\prime})\cdot\bp}\;\eta(\bx +\bq^{\prime})
\overline{\eta(\bx^{\prime}+\bq^{\prime})}g(\bx^{\prime})\;
d\bx^{\prime}\right]\; d\bq^{\prime}\;d\bp\nonumber\\
&&=\int_{\mathbb{R}^{2}\times\mathbb{R}^{2}}\!\!q_{1}^{\prime}\;\left[\int_{\mathbb{R}^{2}}
e^{i (\bx -
\bx^{\prime})\cdot \bp}\eta(\bx +\bq^{\prime})
\overline{\eta(\bx^{\prime}+\bq^{\prime})}g(\bx^{\prime})\;
d\bx^{\prime}\right]\; d\bq^{\prime}d\bp\nonumber\\
&& - \frac{\lambda}{2m^{2}}\int_{\mathbb{R}^{2}\times\mathbb{R}^{2}}\!\!p_{2}
\left[\int_{\mathbb{R}^{2}}
e^{i (\bx-\bx^{\prime})\cdot \bp}\eta(\bx +
\bq^{\prime})\overline{\eta(\bx^{\prime}+
\bq^{\prime})}g(\bx^{\prime})d\bx^{\prime}\right]\;
d\bq^{\prime}\;d\bp\; .
\end{eqnarray}
Let us consider the first integral in (\ref{proofthmseond}). Assuming $\eta$ to be
sufficiently smooth functions, we have
\begin{eqnarray}\label{proofthmthird}
\lefteqn{\int_{\mathbb{R}^{2}\times\mathbb{R}^{2}}q_{1}^{\prime}\;
\left[\int_{\mathbb{R}^{2}}e^{i ({\bx}-{\bx}^{\prime})\cdot
{\bp}}\;\eta({\bx}+ {\bq}^{\prime})
\overline{\eta({\bx}^{\prime}+{\bq}^{\prime})}
g({\bx}^{\prime})d{\bx}^{\prime}\right]\;d{\bq}^{\prime}\;d{\bp}}\nonumber\\
&&=(2\pi)^{2}\int_{\mathbb{R}^{2}}q_{1}^{\prime}
\left[\int_{\mathbb{R}^{2}}
\delta({\bx}- {\bx}^{\prime})]\; \eta({\bx}+
{\bq}^{\prime})\overline{\eta({\bx}^{\prime}+
{\bq}^{\prime})}g({\bx}^{\prime})d{\bx}^{\prime}\right]d{\bq}^{\prime}\nonumber\\
&&=(2\pi)^{2}\int_{\mathbb{R}^{2}}q_{1}^{\prime}
\;|\eta({\bx}+{\bq}^{\prime})|^{2}g({\bx})\;d{\bq}^{\prime}
\end{eqnarray}
Making a second change of variables, ${\bx}+{\bq}^{\prime}=-{\bu}$, the last term in
(\ref{proofthmthird}) becomes
\begin{eqnarray}
\lefteqn{(2\pi)^{2}\int_{\mathbb{R}^{2}}{q_{1}^{\prime} }
|\eta({\bx}+{\bq}^{\prime})|^{2}g({\bx})\;d{\bq}^{\prime}}\nonumber\\
&&=(2\pi)^{2}{x_{1}g(\underbar{x})}\int_{\mathbb{R}^{2}}
|\eta({\bu})|^{2}\; d{u}+(2\pi)^{2}{g(\underbar{x})}
\int_{\mathbb{R}^{2}}u_{1}|\eta({\bu})|^{2}g({\bx})\;d{\bu}\; .
\end{eqnarray}
The second integral in the last line vanishes since, in view of the imposed symmetry, $\eta$ is
an even function of $u_1$.
Thus, noting the normalization of $\eta$ in (\ref{non-comm-CS}),
\be
\int_{\mathbb{R}^{2}\times\mathbb{R}^{2}}q_{1}^{\prime}\;
\left[\int_{\mathbb{R}^{2}}e^{i ({\bx}-{\bx}^{\prime})\cdot
{\bp}}\;\eta({\bx}+ {\bq}^{\prime})
\overline{\eta({\bx}^{\prime}+{\bq}^{\prime})}
g({\bx}^{\prime})d{\bx}^{\prime}\right]\;d{\bq}^{\prime}\;d{\bp} = x_{1}g({\bx}).
\label{proofthmfirstpart}
\en

Next, we observe that,
\beano
\lefteqn{-i\frac{\partial}{\partial x_{2}}\left( e^{i({\bx}-{\bx}^{\prime})\cdot{\bp}}
\;\eta({\bx}+{\bq}^{\prime})\overline{\eta({\bx}^{\prime}+
{\bq}^{\prime})}g({\bx}^{\prime})\right)}\\
&& = p_{2}e^{i ({\bx}- {\bx}^{\prime})\cdot{\bp}}\;
\overline{\eta({\bx}^{\prime}+ {\bq}^{\prime})}g({\bx}^{\prime})
\eta({\bx}+{\bq}^{\prime})\\
 && +  e^{i({\bx}-{\bx}^{\prime})\cdot{\bp}}
\overline{\eta({\bx}^{\prime}+ {\bq}^{\prime})}g({\bx}^{\prime})
\left(-i\frac{\partial}{\partial x_{2}}\right)(\eta({\bx}+{\bq}^{\prime})),
\enano
so that the second integral in  (\ref{proofthmseond}) becomes
\begin{eqnarray}\label{equationtwoparts}
\lefteqn{\frac{\lambda}{2m^{2}}\int_{\mathbb{R}^{2}\times\mathbb{R}^{2}}\!\!p_{2}
\left[\int_{\mathbb{R}^{2}}
e^{i({\bx}-{\bx}^{\prime})\cdot{\bp}}\;
\eta({\bx}+{\bq}^{\prime})\overline{\eta({\bx}^{\prime}+
{\bq}^{\prime})}g({\bx}^{\prime})\;d{\bx}^{\prime}\right]\;{d{\bq}^{\prime}\;d{\bp}}}\nonumber\\
&&=\frac{\lambda}{2m^{2}}\int_{\mathbb{R}^{2}\times\mathbb{R}^{2}}\!\!
\left[\int_{\mathbb{R}^{2}}
\left(-i\frac{\partial}{\partial x_{2}}\right)(e^{i({\bx}-{\bx}^{\prime})\cdot{\bp}}\;
\eta({\bx}+{\bq}^{\prime})\overline{\eta({\bx}^{\prime}+
{\bq}^{\prime})}g({\bx}^{\prime}))\;d{\bx}^{\prime}\right]\;{d{\bq}^{\prime}\;d{\bp}}\nonumber\\
&&-\frac{\lambda}{2m^{2}}\int_{\mathbb{R}^{2}\times\mathbb{R}^{2}}\!\!
\left[\int_{\mathbb{R}^{2}}
\{e^{i({\bx}- {\bx}^{\prime})\cdot {\bp}}\;
\overline{\eta({\bx}^{\prime}+{\bq}^{\prime})}g({\bx}^{\prime})
\left(-i\frac{\partial}{\partial x_{2}}\right)\eta({\bx}+ {\bq}^{\prime})\}\;d{\bx}^{\prime}\right]\;{d{\bq}^{\prime}\;d{\bp}}.\nonumber\\
\end{eqnarray}
Assuming the usual smoothness condition on $\eta$ and again introducing a delta-distribution in $\bx , \bx^\prime$, the first integral on the right hand side  of (\ref{equationtwoparts}) gives
\begin{eqnarray}\label{secndsummand}
\lefteqn{\frac{\lambda}{2m^{2}}\int_{\mathbb{R}^{2}\times\mathbb{R}^{2}}
\!\!\left(-i\frac{\partial}{\partial x_{2}}\right)\left[\left\{\int_{\mathbb{R}^{2}}
e^{i({\bx}-{\bx}^{\prime})\cdot {\bp}}\;\eta({\bx}+
{\bq}^{\prime})\overline{\eta({\bx}^{\prime}+{\bq}^{\prime})}
g({\bx}^{\prime})d{\bx}^{\prime}\right\}\;d{\bq}^{\prime}\;d{\bp}\right]}\nonumber\\
&&=\frac{(2\pi)^{2}\lambda}{2m^{2}}
\int_{\mathbb{R}^{2}}\!\!\left(-i\frac{\partial}{\partial x_{2}}\right)\left[\left\{\int_{\mathbb{R}^{2}}\!\!\!
\delta({\bx}-{\bx}^{\prime})\eta({\bx}+{\bq}^{\prime})
\overline{\eta({\bx}^{\prime}+{\bq}^{\prime})}g({\bx}^{\prime})
d{\bx}^{\prime}\right\}\;d{\bq}^{\prime}\right]\nonumber\\
&&=-\frac{i\lambda}{2m^{2}}\frac{\partial}{\partial x_{2}}g({\bx})\; .
\end{eqnarray}
Similarly, the second integral (\ref{equationtwoparts}) yields
\beano
\lefteqn{-\frac{\lambda}{2m^{2}}\int_{\mathbb{R}^{2}\times\mathbb{R}^{2}}\!\!
\left[\int_{\mathbb{R}^{2}}
\{e^{i({\bx}-{\bx}^{\prime})\cdot\underbar{p}}\;
\overline{\eta({\bx}^{\prime}+{\bq}^{\prime})}g({\bx}^{\prime})
\left(-i\frac{\partial}{\partial x_{2}}\right)\eta({\bx}+{\bq}^{\prime})\}\;
d{\bx}^{\prime}\right]\;{d{\bq}^{\prime}\;d{\bp}}}\\
&&=-\frac{\lambda}{2m^{2}}\int_{\mathbb{R}^{2}}
\left[\int_{\mathbb{R}^{2}}\{\delta({\bx}-{\bx}^{\prime})
\overline{\eta({\bx}^{\prime}+{\bq}^{\prime})}g({\bx}^{\prime})
\left(-i\frac{\partial}{\partial x_{2}}\right)\eta({\bx}+ {\bq}^{\prime})\}\;d\b{x}^{\prime}\}\right]
\;d\underbar{q}^{\prime}\\
&&=-\frac{\lambda}{2m^{2}}g({\bx})
\int_{\mathbb{R}^{2}}\overline{\eta(\underbar{x}+
{\bq}^{\prime})}\left(-i\frac{\partial}{\partial x_{2}}\right)\eta({\bx}+{\bq}^{\prime})\;d{\bq}^{\prime}.
\enano
Introducing another change of variables,  ${\bx}+{\bq}^{\prime}={\bu}$, this
becomes
\bea\label{finalzeroterm}
\lefteqn{-\frac{\lambda}{2m^{2}}g({\bx})
\int_{\mathbb{R}^{2}}\overline{\eta({\bu})}\left(-i\frac{\partial}{\partial u_{2}}\right)\eta({\bu})\;d{\bu}}\nonumber\\
&& =\frac{i\lambda}{2m^{2}}g({\bx})
\int_{\mathbb{R}^{2}}\overline{\eta({\bu})}\frac{\partial}{\partial u_{2}}\eta({\bu})\;d{\bu} = 0,
\ena
the last equality following since, in view of the evenness of $\eta$, the derivative term,
$\dfrac{\partial}{\partial u_{2}}\eta({\bu})$, is an odd function.

Thus finally, combining   (\ref{finalzeroterm}) with (\ref{proofthmseond}), (\ref{proofthmfirstpart}), and (\ref{secndsummand}), we obtain
\begin{equation*}
(\hat{\mathcal{O}}_{q_{1}}g)({\bx})=\left(x_{1}-\frac{i\lambda}{2m^{2}}
\frac{\partial}{\partial x_{2}}\right)g({\bx}).
\end{equation*} \qed

\bigskip
\prf {\bf of Theorem \ref{ineqvlocalexpgrptrns}.}

Using (\ref{defcentrlextfirst}) and (\ref{defcentrlexsecond}), it can easily be verified that $\xi$, $\xi^{\prime}$, and $\xi^{\prime\prime}$ given in Proposition (\ref{ineqvlocalexpgrptrns}) are local exponents for the group of translations $G_{T}$ in $\mathbb{R}^{4}$.

It remains to prove the inequivalence of the given multipliers. Let us first prove the fact that $\xi_{1}:=\xi-\xi^{\prime}$ is not trivial. Indeed we have,
\begin{eqnarray}
\lefteqn{\xi_{1}((q_{1},q_{2},p_{1},p_{2}),(q_{1}^{\prime},q_{2}^{\prime},p_{1}^{\prime},p_{2}^{\prime}))}\nonumber\\
&&=\xi((q_{1},q_{2},p_{1},p_{2}),(q_{1}^{\prime},q_{2}^{\prime},p_{1}^{\prime},p_{2}^{\prime}))-\xi^{\prime}((q_{1},q_{2},p_{1},p_{2}),(q_{1}^{\prime},q_{2}^{\prime},p_{1}^{\prime},p_{2}^{\prime}))\nonumber\\
&&=\frac{1}{2}q_{1}p_{1}^{\prime}+\frac{1}{2}q_{2}p_{2}^{\prime}+\frac{1}{2}p_{2}p_{1}^{\prime}-\frac{1}{2}p_{1}q_{1}^{\prime}-\frac{1}{2}p_{2}q_{2}^{\prime}-\frac{1}{2}p_{1}p_{2}^{\prime}.
\end{eqnarray}
Now from (\ref{deftrivcentrlex}), it follows immediately that triviality of a multiplier $\eta$ for some abelian group in terms of a suitable continous function implies the fact that $\eta(g,g^{\prime})=\eta(g^{\prime},g)$ holds for any two group elements of the given abelian group. By contrapositivity, $\eta(g,g^{\prime})\neq\eta(g^{\prime},g)$ guarantees the nontriviality of the multiplier in question.

In other words, to prove the nontriviality of $\xi_{1}$, it suffices to show that $$\xi_{1}((q_{1},q_{2},p_{1},p_{2}),(q_{1}^{\prime},q_{2}^{\prime},p_{1}^{\prime},p_{2}^{\prime}))
\neq\xi_{1}((q_{1}^{\prime},q_{2}^{\prime},p_{1}^{\prime},p_{2}^{\prime}),
(q_{1},q_{2},p_{1},p_{2}))$$ always holds. Indeed,
\begin{eqnarray}
\lefteqn{\xi_{1}((q_{1}^{\prime},q_{2}^{\prime},p_{1}^{\prime},p_{2}^{\prime}),(q_{1},q_{2},p_{1},p_{2}))}\nonumber\\
&&=\frac{1}{2}q_{1}^{\prime}+\frac{1}{2}q_{2}^{\prime}p_{2}+\frac{1}{2}p_{2}^{\prime}p_{1}-\frac{1}{2}p_{1}^{\prime}q_{1}-\frac{1}{2}p_{2}^{\prime}q_{2}-\frac{1}{2}p_{1}^{\prime}p_{2},\nonumber\\
&&=-\xi_{1}((q_{1},q_{2},p_{1},p_{2}),(q_{1}^{\prime},q_{2}^{\prime},p_{1}^{\prime},p_{2}^{\prime})).
\end{eqnarray}
Let us now prove that $\xi_{2}:=\xi^{\prime}-\xi^{\prime\prime}$ is nontrivial. We have,
\begin{eqnarray}
\lefteqn{\xi_{2}((q_{1},q_{2},p_{1},p_{2}),(q_{1}^{\prime},q_{2}^{\prime},p_{1}^{\prime},p_{2}^{\prime}))}\nonumber\\
&&=\xi^{\prime}((q_{1},q_{2},p_{1},p_{2}),(q_{1}^{\prime},q_{2}^{\prime},p_{1}^{\prime},p_{2}^{\prime}))-\xi^{\prime\prime}((q_{1},q_{2},p_{1},p_{2}),(q_{1}^{\prime},q_{2}^{\prime},p_{1}^{\prime},p_{2}^{\prime}))\nonumber\\
&&=\frac{1}{2}[p_{1}p_{2}^{\prime}+q_{1}q_{2}^{\prime}-p_{2}p_{1}^{\prime}-q_{2}q_{1}^{\prime}]\nonumber\\
&&=-\xi_{2}((q_{1}^{\prime},q_{2}^{\prime},p_{1}^{\prime},p_{2}^{\prime}),(q_{1},q_{2},p_{1},p_{2})).
\end{eqnarray}
The above equation reflects the fact that $\xi_{2}$ is indeed nontrivial which in turn implies that $\xi^{\prime}$ and $\xi^{\prime\prime}$ are inequivalent.
Hence it follows that $\xi$, $\xi^{\prime}$ and $\xi^{\prime\prime}$ are three inequivalent local exponents of $G_{T}$.
\qed


\end{document}